\documentclass[%
 reprint,
preprintnumbers,
nofootinbib,
 amsmath,amssymb,
 aps,
 prd,
]{revtex4-2}

\usepackage{graphicx}
\usepackage{dcolumn}
\usepackage{bm}
\usepackage{hyperref}
\usepackage{xcolor}
\usepackage{natbib}
\usepackage{xspace}
\usepackage{soul}
\usepackage[capitalize]{cleveref}
\usepackage{subcaption}
\usepackage[normalem]{ulem}
\usepackage{physics}

\renewcommand{\Re}{\text{Re}}

\newcommand{\BE}{\begin{equation}}
\newcommand{\EE}{\end{equation}}

\newcommand{\nn}{\nonumber}
\newcommand{\ord}[1]{{\ensuremath \mathcal{O}(#1)}}

\newcommand{\mhp}[1]{m_{H^\pm}^{#1}}

\newcommand{\brr}{\ensuremath{R[\text{BR}(h\to\gamma\gamma)]}}
\newcommand{\gammar}{\ensuremath{R[\Gamma(h\to\gamma\gamma)]}}

\newcommand{\decoup}{DI}

\begin{document}

\preprint{DESY-23-104}
\preprint{KEK-TH-2539}
\preprint{OU-HET-1194}

\title{
Leading two-loop corrections to the Higgs di-photon decay in the Inert Doublet Model}
\author{Masashi Aiko$^{1}$}
\email{m-aiko@miyakonojo.kosen-ac.jp}
\author{Johannes Braathen$^2$}
\email{johannes.braathen@desy.de}
\author{Shinya Kanemura$^3$}
\email{kanemu@het.phys.sci.osaka-u.ac.jp}
\affiliation{$^1$ National Institute of Technology, Miyakonojo College, Miyakonojo, Miyazaki 885-8567, Japan}
\affiliation{$^2$ Deutsches Elektronen-Synchrotron DESY, Notkestr.~85, 22607 Hamburg, Germany\\}
\affiliation{$^3$ Department of Physics, Osaka University, Toyonaka, Osaka 560-0043, Japan}
\date{\today}

\begin{abstract}
    Leading two-loop contributions to the di-photon decay of the Higgs boson are evaluated for the first time in the Inert Doublet Model (IDM). We employ for this calculation the Higgs low-energy theorem, meaning that we obtain corrections to the Higgs decay process by taking Higgs-field derivatives of the leading two-loop contributions to the photon self-energy. Specifically, we have included corrections involving inert BSM Higgs bosons and gauge bosons, as well as external-leg contributions involving inert scalars, gauge bosons and fermions. Our calculation has been performed with a full on-shell renormalization, and in the gauge-less limit. Moreover, we performed two independent calculations, using the background-field method and the pinch technique, in order to apply the Higgs low-energy theorem consistently, and found full agreement between the two calculations. We investigate our results numerically in two scenarios of the IDM: one with a light dark matter (DM) candidate (Higgs resonance scenario), and another with all additional scalars heavy (heavy Higgs scenario). In both cases, we find that the inclusion of two-loop corrections qualitatively modifies the behavior of the decay width, compared with the one-loop ($i.e.$ leading) order, and that they increase the deviation from the Standard Model. Furthermore, we demonstrate that the inclusion of the newly-computed two-loop corrections is essential to reliably interpret the observation or non-observation of a deviation in the Higgs di-photon decay width at current and future colliders.  
\end{abstract}

\maketitle


\section{Introduction}
In spite of its successes, the Standard Model (SM) of Particle Physics cannot explain the existence of dark matter (DM), the observed tiny neutrino masses, or the baryon asymmetry of the universe. Therefore, the SM must be extended to address these problems. At the same time, while the Higgs sector has now been confirmed as the origin of the electroweak symmetry breaking, there is no unequivocal guiding principle for its building, and its structure remains uncharted. Moreover, because many issues left unanswered by the SM can be related to the Higgs boson, Beyond-the-Standard-Model (BSM) theories commonly feature extended scalar sectors. 

Although the contribution of DM to the energy budget of the Universe ($i.e.$ the DM relic density) has been measured to a high level of precision by the PLANCK collaboration~\cite{Planck:2018vyg}, its nature  is, to this day, still a mystery. Among the many possible explanations of DM, the scenario of it being a weakly interacting massive particle (WIMP),  remains promising and motivated due to the DM candidate being around the electroweak scale and therefore the testability of this scenario. An interesting possibility in this context is to explain the nature of DM via an extension of the Higgs sector. 

The inert doublet model (IDM)~\cite{Deshpande:1977rw, Barbieri:2006dq} is a simple extension of the SM including a DM candidate. In this model, an additional $SU(2)_{L}$ doublet field and an unbroken $Z_{2}$ symmetry are introduced. Under the $Z_{2}$ symmetry, all of the SM fields are even, while the additional doublet field is odd. Therefore, the lightest neutral component of the additional doublet is stable and it can be a DM candidate. Since the $Z_{2}$-odd particles (also called inert particles) interact with the SM particles through gauge interactions and scalar self-interactions, they are thermalized in the early universe. Thus, the dark matter abundance in the present universe can be generated as a thermal relic. Moreover, extending the IDM by including right-handed neutrinos allows explaining the observed tiny neutrino masses via the radiative seesaw mechanism --- this model is referred to as the Tao-Ma model~\cite{Tao:1996vb,Ma:2006km}.

Direct detection experiments of DM, such as LUX-ZEPLIN (LZ)~\cite{LZ:2022ufs,LZ:2024} and XENONnT~\cite{XENON:2023sxq} constrain the allowed mass range of the DM within the IDM. In addition, the IDM has been tested by collider experiments through searches for the direct production of the $Z_{2}$-odd scalars~\cite{Pierce:2007ut, Lundstrom:2008ai} and electroweak precision tests~\cite{ParticleDataGroup:2022pth}. On the other hand, the decays of the discovered Higgs boson also constrain the parameter space of the IDM. The Higgs boson decays into an invisible DM pair constrains the size of Higgs-DM coupling if the decay channel is kinematically allowed. In addition, radiative effects from the $Z_{2}$-odd particles modify the partial decay widths of the Higgs boson with respect to those in the SM~\cite{Arhrib:2015hoa, Kanemura:2016sos, Kanemura:2019kjg}. These signals can be tested by precision measurements of the Higgs boson decays at the High-Luminosity LHC (HL-LHC)~\cite{Apollinari:2017lan} and at future lepton colliders, such as the International Linear Collider (ILC)~\cite{ILC:2013jhg}, the Future Circular Collider~\cite{FCC:2018evy}, or the Circular Electron Positron Collider~\cite{CEPCStudyGroup:2018ghi}.

Several options can be explored to accommodate DM in the IDM~\cite{Kanemura:2016sos}, either with a light DM candidate --- with a mass below or around half of the Higgs mass --- or with a relatively heavy one, at a few hundred GeV or more. These scenarios often involve significant mass splittings between the DM candidate and the other $\mathbb{Z}_2$-odd scalars --- this is especially the case when DM is light. In turn, such mass splittings typically give rise to large BSM deviations in couplings or decay widths of the 125-GeV Higgs boson, due to non-decoupling effects in radiative corrections involving the heavy BSM scalars. This was first pointed out for the trilinear Higgs coupling~\cite{Kanemura:2002vm,Kanemura:2004mg}, for which it is known that deviations of $\mathcal{O}(100\%)$ from the SM are possible in various models with extended scalar sectors~\cite{Aoki:2012jj,Kanemura:2015mxa,Kanemura:2015fra,Arhrib:2015hoa,He:2016sqr,Kanemura:2016lkz,Kanemura:2016sos,Kanemura:2017wtm,Kanemura:2017wtm,Senaha:2018xek,Braathen:2019pxr,Braathen:2019zoh,Braathen:2020vwo,Bahl:2022jnx}. For other Higgs couplings, $e.g.$ to gauge bosons or fermions, effects from mass splittings are usually less dramatic~\cite{Kanemura:2015mxa,Kanemura:2019kjg}, but at the same time these couplings are much better constrained than the trilinear Higgs coupling (see for instance Refs.~\cite{ATLAS:2022vkf,CMS:2022dwd}). In this context, a property of the Higgs boson that is especially useful to investigate the parameter space of the IDM is its \emph{di-photon decay width}, because of the existence of a charged scalar boson in this model. The one-loop BSM contributions to this decay have been studied in Refs.~\cite{Arhrib:2012ia, Swiezewska:2012eh, Kanemura:2016sos} and, in particular, it has been found that the deviation in the effective Higgs coupling with photons can reach about $-5\%$ in scenarios with light DM~\cite{Kanemura:2016sos}. The magnitude of this deviation is thus comparable with the expected precision --- of about 7\% at the 95\% confidence level (CL) --- at the HL-LHC~\cite{Cepeda:2019klc}, and this strongly motivates studying the impact of the two-loop --- or in other words the next-to-leading order (NLO) --- corrections to the di-photon decay width. 

In this paper, we therefore evaluate the leading two-loop contributions to the di-photon decay width of the Higgs boson in the IDM. We employ the Higgs low-energy theorem (LET)~\cite{Ellis:1975ap, Shifman:1979eb, Kniehl:1995tn} for the calculation of the leading two-loop corrections. Following this theorem, an effective Higgs-photon coupling is obtained from the photon self-energy by taking a Higgs-field derivative. Because we are investigating BSM effects related to (potentially large) scalar quartic couplings, we perform an expansion in powers of the comparatively much smaller EW gauge couplings, and retain only the leading terms of this expansion. At the same time, the ratio of the two EW gauge couplings, related to the weak mixing angle, is kept fixed (see $e.g.$ Ref.~\cite{Hessenberger:2016atw}). We have included in our calculation two-loop corrections involving inert scalars, as well as external-leg contributions involving the inert scalars, gauge bosons, and fermions are included. We have employed an on-shell renormalization scheme --- see in particular Refs.~\cite{Kanemura:2016sos,Braathen:2019pxr,Braathen:2019zoh}. Recently, leading two-loop corrections to the Higgs di-photon decay have also been investigated in a scenario of the 2HDM with alignment in the Higgs sector~\cite{Degrassi:2023eii} and in a real triplet model~\cite{Degrassi:2024qsf}. In contrast to these articles, we consider here a new and different model --- namely the IDM --- and we consider our results in light of DM phenomenology. Specifically, we investigate our results numerically in two scenarios of the IDM: one with a light DM candidate, and another with all additional scalars heavy. In both cases, we find that the inclusion of two-loop corrections qualitatively modifies the behavior of the decay width, compared with the one-loop order, and that they increase the deviation from the SM. We also study the correlation between the Higgs partial decay width to two photons and the trilinear Higgs coupling, and how this is modified by the inclusion of two-loop corrections to both quantities. Our work can furthermore be distinguished by the computational techniques that we employ: indeed, we perform our calculations using the background field method as well as the pinch technique, while Refs.~\cite{Degrassi:2023eii,Degrassi:2024qsf} employ the unitary gauge. 

This article is organized as follows: we define in \cref{sec:IDM} our notations for the IDM and the considered theoretical and experimental constraints. In \cref{sec:calc}, we present our setup for the calculation of the leading two-loop corrections to the Higgs di-photon decay and our analytical results. Numerical investigations of our results are shown in \cref{sec:numres}, followed by a discussion of their implications and our conclusions in \cref{sec:ccl}. Different appendices contain additional analytical expressions, complementing the results presented in \cref{sec:calc}: in \cref{app:2L_res_IDM_decoup}, we provide results for the two-loop contributions to the Higgs decay width to two photons using the decoupling-inspired scheme of Refs.~\cite{Braathen:2019pxr,Braathen:2019zoh} for the BSM mass parameter $\mu_2$; in \cref{app:2Lphotonself}, we present our analytical expressions for the unrenormalized two-loop corrections to the photon self-energy; and finally, in \cref{app:2L_hhh}, we give results for two-loop corrections for the trilinear Higgs coupling applicable in any scenario of the IDM (extending the results of Refs.~\cite{Braathen:2019pxr,Braathen:2019zoh}). 

\section{The Inert Doublet Model}
\label{sec:IDM}
The Higgs sector of the IDM is composed of two isospin doublet scalar fields $\Phi_{1}$ and $\Phi_{2}$ with an unbroken $Z_{2}$ symmetry.
The inert doublet field $\Phi_{2}$ is $Z_{2}$-odd, while all the other fields are $Z_{2}$-even.
The tree-level Higgs potential is given by
\begin{align}
V &= \mu_{1}^{2}\abs{\Phi_{1}}^{2}+\mu_{2}^{2}\abs{\Phi_{2}}^{2}
+\frac{\lambda_{1}}{2}\abs{\Phi_{1}}^{4}+\frac{\lambda_{2}}{2}\abs{\Phi_{2}}^{4}  \\
&\quad
+\lambda_{3}\abs{\Phi_{1}}^{2}\abs{\Phi_{2}}^{2}+\lambda_{4}\abs{\Phi_{1}^{\dagger}\Phi_{2}}^{2}+\frac{\lambda_{5}}{2}\qty[(\Phi_{1}^{\dagger}\Phi_{2})^{2}+h.c.]\,.\nn
\label{eq: Higgs_potential}
\end{align}
The phase of $\lambda_{5}$ can be removed via a global phase rotation, $\Phi_{2}\to e^{i\theta}\Phi_{2}$, without affecting any other part of the Lagrangian.
Therefore, we take $\lambda_{5}$ to be real and negative.

In the inert vacuum phase, where only $\Phi_{1}$ acquires a vacuum expectation value (VEV) $v(\simeq 246\text{ GeV})$, the scalar doublet fields are parameterized as
\begin{align}
\Phi_{1} = \mqty(G^{+} \\ \frac{1}{\sqrt{2}}(v+h+iG^{0}))\qc
\Phi_{2} = \mqty(H^{+} \\ \frac{1}{\sqrt{2}}(H+iA)),
\end{align}
where $h$ is the discovered Higgs boson with a mass of 125-GeV and $G^{\pm}$ and $G^{0}$ are (would-be) Nambu-Goldstone (NG) bosons.
We call the additional $Z_{2}$-odd scalar bosons $H,\, A$, and $H^{\pm}$ the inert scalar bosons.

The Higgs-field dependent masses of the scalar bosons are given by
\begin{align}
\label{eq:0L_masses}
m_{h}^{2}(h) &= \mu_1^2+\frac{3}{2}\lambda_{1}(v+h)^{2}\,, \nn\\
m_G^2(h)& =\mu_1^2+\frac{1}{2}\lambda_1(v+h)^2+m_Z^2\,,\nn\\
m_{G^\pm}^2(h)& =\mu_1^2+\frac{1}{2}\lambda_1(v+h)^2+m_W^2\,,\nn\\
m_{H^{\pm}}^{2}(h) &= \mu_{2}^{2}+\frac{1}{2}\lambda_{3}(v+h)^{2}\,,\nn \\
m_{H}^{2}(h) &= m_{H^{\pm}}^{2}(h)+\frac{1}{2}(\lambda_{4}+\lambda_{5})(v+h)^{2}\,,\nn \\
m_{A}^{2}(h) &= m_{H^{\pm}}^{2}(h)+\frac{1}{2}(\lambda_{4}-\lambda_{5})(v+h)^{2}\,,
\end{align}
and can be simplified at the minimum of the potential, $h=0$, using the stationary (or minimization) condition $\mu_1^2=-\frac{1}{2}\lambda_1 v^2$. 
Since we take $\lambda_{5}\le 0$, $H$ is lighter than $A$ and becomes a DM candidate. We employ a standard Feynman gauge fixing, expressed only in terms of the Higgs VEV but of not the Higgs field, meaning that $m_W^2$ and $m_Z^2$ are here the field-independent masses of the $W$ and $Z$ gauge bosons. This treatment is compatible with the application of the Higgs LET~\cite{Pilaftsis:1997fe}. 

There are theoretical constraints on the Higgs potential parameters such as vacuum stability, perturbative unitarity, and the inert vacuum condition. Vacuum stability bounds require that the Higgs potential is bounded from below in any direction of field space with large field values. The necessary conditions for vacuum stability are given by~\cite{Deshpande:1977rw, Kanemura:1999xf}
\begin{align}
&\lambda_{1}>0\qc
\lambda_{2}>0\,,\nn\\
&\sqrt{\lambda_{1}\lambda_{2}}+\lambda_{3}+\mathrm{MIN}(0, \lambda_{4}+\lambda_{5}, \lambda_{4}-\lambda_{5})>0\,.\label{eq:vac_stab_2}
\end{align}
Perturbative unitarity bounds impose $\abs{a_{0}^{i}}\leq 1/2$, where $a_{0}^{i}$ are the eigenvalues of the $s$-wave amplitude matrix. In the high-energy limit, only two-to-two elastic scatterings of scalar bosons are relevant, and explicit formulae for $a_{0}^{i}$ are given in Refs.~\cite{Kanemura:1993hm, Akeroyd:2000wc}. The inert vacuum condition requires the following inequality~\cite{Ginzburg:2010wa},
\begin{align}
\frac{\mu_{1}^{2}}{\sqrt{\lambda_{1}}} < \frac{\mu_{2}^{2}}{\sqrt{\lambda_{2}}}.
\end{align}
From the vacuum stability conditions, $\lambda_{1}$ and $\lambda_{2}$ are positive, while $\mu_{1}^{2}$ is negative due to the stationary condition. Therefore, $\mu_{2}^{2}>0$ is a sufficient condition to realize a stable inert vacuum. As a last theoretical constraint, we require that all scalar quartic couplings fulfill the inequality $|\lambda_i|\leq4\pi$, as a criterion to ensure perturbativity.

Direct collider searches at LEP as well as measurements of the electroweak precision observables (EWPOs) give bounds on the masses of the inert scalar bosons. The measurements of the $Z$ and $W^{\pm}$ bosons' widths lead to the lower limits on the inert scalar masses
\begin{align}
&m_{H}+m_{A} \ge m_{Z}\qc
2m_{H^{\pm}} \ge m_{Z},\nn\\
&m_{H, A}+m_{H^{\pm}} \ge m_{W},
\end{align}
in order to kinematically prohibit the decay processes $Z\to HA,\, H^{+}H^{-}$ and $W^{\pm}\to HH^{\pm},\, AH^{\pm}$. From searches of the $e^{+}e^{-}\to H^{+}H^{-}$ production process, we have~\cite{Pierce:2007ut}
\begin{align}
m_{H^{\pm}}\gtrsim 90\, \mathrm{GeV}.
\end{align}
On the other hand, from the $e^{+}e^{-}\to HA$ production process, we have~\cite{Lundstrom:2008ai}
\begin{align}
m_{H} > 80\, \mathrm{GeV}\qq{or}
m_{A} > 100\, \mathrm{GeV}.
\end{align}
If the mass difference between $A$ and $H$ is smaller than 8 GeV, there remain allowed regions in the mass range below $80-100$ GeV. Current bounds from searches of inert scalars at the LHC have been discussed for instance in Refs.~\cite{Belanger:2015kga,Ilnicka:2015jba,Belyaev:2018ext,Kalinowski:2020rmb}, but do not produce significantly more stringent limits than LEP searches at the moment. 

The inert scalar contributions to the EWPOs can be parameterized by the electroweak parameters $\Delta S,\, \Delta T$, and $\Delta U$~\cite{Peskin:1991sw}. One-loop analytic expressions for these are given in Ref.~\cite{Kanemura:2011sj, Hessenberger:2016atw}. From a global fit analysis, the constraints on the $\Delta S$ and $\Delta T$ parameters are given by~\cite{ParticleDataGroup:2022pth}
\begin{align}
\label{eq:DeltaSDeltaT}
\Delta S = -0.01\pm 0.07\qc
\Delta T = 0.04\pm 0.06,
\end{align}
when fixing $\Delta U=0$. The correlation coefficient in the $\chi^{2}$ analysis is $+0.92$. We require $\Delta S$ and $\Delta T$ to be within the $95\%$ CL intervals of the values given in \cref{eq:DeltaSDeltaT}. We note that as an additional check, we have also verified for our numerical benchmark scenarios (discussed in the following), that the electroweak precision observables at the $Z$ pole --- $i.e.$ the $W$-boson mass, the sine of the effective weak mixing angle, and the $Z$-boson decay width --- evaluated in the IDM at one and two loops with \texttt{THDM\_EWPOS}~\cite{Hessenberger:2016atw,Hessenberger:2022tcx} are all within 95\% CL from their experimentally measured values.

Experimental limits~\cite{ATLAS:2022vkf,CMS:2022dwd,ATLAS:2023tkt,CMS:2023sdw} on the branching ratio of invisible decays of the detected Higgs boson provide a complementary probe of models with inert scalars like the IDM. The corresponding constraints depend on the chosen scenario and will be verified for the benchmark scenarios considered in \cref{sec:numres}. 

The last experimental constraints that we take into account relate to DM phenomenology. The DM relic density has been determined as $\Omega_{\rm DM} h^{2}=0.1200\pm0.0012$ from PLANCK data~\cite{Planck:2018vyg}. Using the code \texttt{micrOMEGAs}~\cite{Belanger:2018ccd,Alguero:2023zol}, we compute $\Omega_{\rm DM} h^{2}$ and exclude the parameter points with an overabundance of DM. We also evaluate the spin-independent cross section of DM scattering by using \texttt{micrOMEGAs} and impose the constraint from DM direct detection by the LZ experiment~\cite{LZ:2024}.

\section{Higgs Low-Energy Theorem and di-photon decay of the Higgs boson}
\label{sec:calc}
We present in this section our calculation of the leading two-loop corrections to $\Gamma(h\to\gamma\gamma)$ with the Higgs LET~\cite{Ellis:1975ap, Shifman:1979eb,Kniehl:1995tn}. Similarly to effective-potential computations (see examples at two loops for the Higgs mass in Ref.~\cite{Slavich:2020zjv} and references therein, or Refs.~\cite{Senaha:2018xek,Braathen:2019pxr,Braathen:2019zoh,Braathen:2020vwo} for the trilinear Higgs coupling), the use of the Higgs LET implies that we neglect the incoming momentum on the Higgs-boson leg (the validity of this approximation will be discussed below). In this case, we can write the amputated amplitude for the di-photon decay as
\begin{align}
 \Gamma^{\mu\nu}(h\to\gamma\gamma)=g^{\mu\nu}\Gamma_{h\to\gamma\gamma}^{(g)}-p_{1}^\nu p_{2}^\mu \Gamma_{h\to\gamma\gamma}^{(p)}\,,
\end{align}
where $g^{\mu\nu}$ is the metric tensor and $p_{1, 2}$ are the photon momenta. The above equation also serves as the definition of the form factors $\Gamma_{h\to\gamma\gamma}^{(g)}$ and $\Gamma_{h\to\gamma\gamma}^{(p)}$. Due to the Ward-Takahashi identity of QED, we obtain $\Gamma_{h\to\gamma\gamma}^{(p)}=\Gamma_{h\to\gamma\gamma}^{(g)}/(p_1\cdot p_2)$, and the decay width is given by
\begin{align}
 \Gamma(h\to\gamma\gamma)=\frac{1}{16\pi m_h}|\Gamma_{h\to\gamma\gamma}^{(g)}|^2\,.
\end{align}
This form factor can be expressed as $\Gamma^{(g)}_{h\to\gamma\gamma}=\ m_h^2C_{h\gamma\gamma}/2$ where $C_{h\gamma\gamma}$ is an effective Higgs-photon coupling constant defined as
\begin{align}
 \mathcal{L}_\text{eff}\supset -\frac{1}{4}C_{h\gamma\gamma}h F_{\mu\nu}F^{\mu\nu}\,,
\end{align}
and which the LET of Ref.~\cite{Kniehl:1995tn} allows us to compute 
simply by taking a Higgs-field derivative of the photon self-energy, $i.e.$
\begin{align}
 C_{h\gamma\gamma}=\frac{\partial}{\partial h}\Pi_{\gamma\gamma}(p^2=0)\bigg|_{h=0}\,.
\end{align}
For this equation, we have expanded the photon self-energy using the Ward-Takahashi identity of QED as 
\begin{align}
 \Sigma_{\gamma\gamma}^{\mu\nu}(p^2)=\big(-p^2g^{\mu\nu}+p^\mu p^\nu\big)\Pi_{\gamma\gamma}(p^2)\,.
\end{align}
We remark that for models such as the IDM, where the 125-GeV Higgs field is aligned in field space with the EW VEV, the derivative with respect to the Higgs field $h$ can be replaced by a derivative with respect to the VEV $v$ --- because field-dependent masses and couplings are always functions of the quantity $v+h$. This is, however, not directly the case for general scenarios (e.g.\ a Two-Higgs-Doublet Model away from the alignment limit).

From this effective coupling, we obtain the Higgs decay width to two photons as 
\begin{align}
 \Gamma(h\to\gamma\gamma)=&\ \frac{m_h^3}{64\pi}|C_{h\gamma\gamma}|^2\,.
\end{align}

\begin{figure}[ht]
\centering
    \centering
    \includegraphics[width=.5\textwidth]{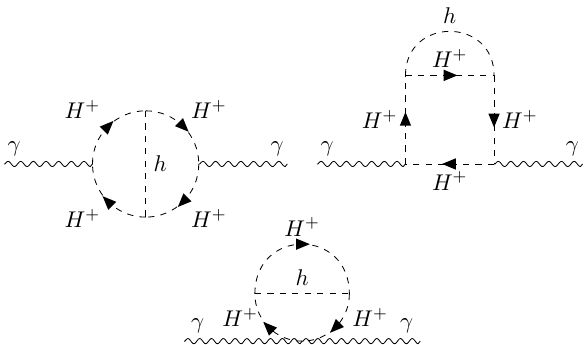}
    \caption{Two-loop diagrams contributing to the photon self-energy at $\order{e^2\lambda_3^2}$.}
   \label{fig:orderlam3sq}
   \vspace{0.5cm}
\end{figure}

\begin{figure}[ht]
    \centering
    \includegraphics[width=.5\textwidth]{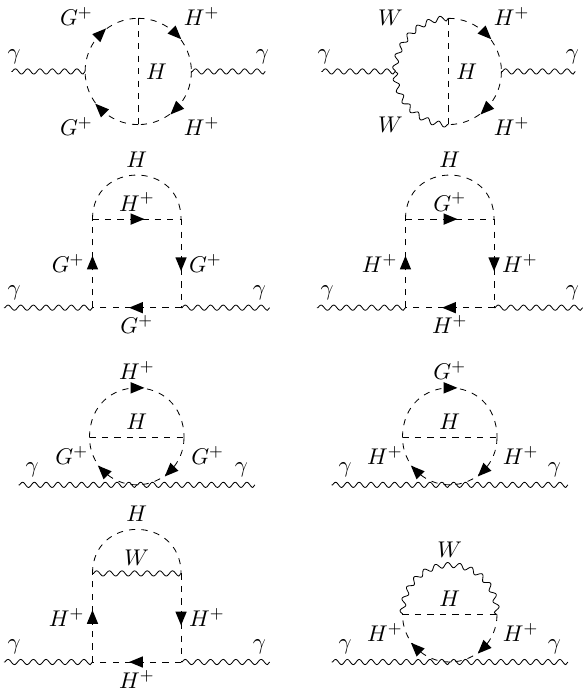}
    \caption{Two-loop diagrams contributing to the photon self-energy at 
    $\order{e^2(\lambda_4+\lambda_5)^2}$. }
    \label{fig:orderlam4+5sq}
    \vspace{0.5cm}
\end{figure}

\begin{figure}[ht]
    \centering    \includegraphics[width=0.375\textwidth]{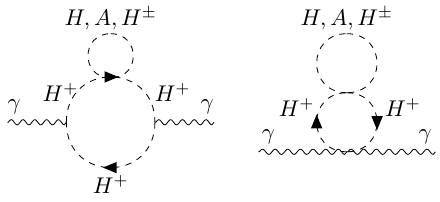}
    \caption{Two-loop diagrams contributing to the photon self-energy at $\order{e^2\lambda_2}$. }
    \label{fig:orderlam2}
\end{figure}

Before turning to the specific steps of our calculation and to our analytical results, some assumptions made in this work should be discussed. We are interested here in the dominant two-loop order (NLO) contributions to the di-photon decay. In terms of couplings, these are corrections involving two powers of the electric charge $e$ and one or two powers of quartic scalar couplings $\lambda_i$. Such contributions arise from \textit{i)} purely-scalar corrections involving the BSM scalars $H$, $A$, and $H^\pm$, as well as \textit{ii)} diagrams involving both BSM scalars and gauge bosons. However, these types of diagrams also yield terms of higher orders in powers of $e$, which are subleading. For this reason, we perform a series expansion of the expressions of the various contributions in the EW gauge couplings around zero, and we only retain terms of quadratic order at most --- $i.e.$ we keep a prefactor $e^2$ in all photon self-energy contributions and their derivatives. We retain, moreover, the dependence on the weak mixing angle (in other words, we take the EW gauge couplings $g_1,\ g_2$ to zero while keeping their ratio $g_2/g_1$ fixed). Furthermore, we neglect throughout this work the quartic coupling $\lambda_1(\simeq 0.25)$, which is related to the mass of the detected Higgs boson, in comparison to the other Lagrangian quartic couplings that can take significantly larger values. Consequently, the field-dependent Higgs and would-be NG boson masses in \cref{eq:0L_masses} simply reduce to $m_h(h)=0$, $m_G(h)=m_Z$ and $m_{G^\pm}(h)=m_W$, and together with the expansion performed in powers of the EW gauge couplings, this corresponds to $m_h,\,m_G,\,m_{G^\pm}\ll m_H,\,m_A,\,m_{H^\pm}$ (although in one of the numerical scenarios we also consider below, we additionally assume $m_H$ to be small). This assumption motivates the use of the Higgs LET, which as mentioned earlier implies that we are neglecting the external momentum $p^2=m_h^2$ on the Higgs leg. At two loops, the genuine corrections can be distinguished into four categories, involving the couplings $\lambda_3^2$, $(\lambda_4+\lambda_5)^2$, $(\lambda_4-\lambda_5)^2$, and $\lambda_2$ respectively. Some example photon self-energy diagrams of orders $\lambda_3^2$, $(\lambda_4+\lambda_5)^2$ and $\lambda_2$ are shown in figures~\ref{fig:orderlam3sq},~\ref{fig:orderlam4+5sq} and~\ref{fig:orderlam2} --- and we note that terms of order $(\lambda_4-\lambda_5)^2$ can straightforwardly be obtained from those of order $(\lambda_4+\lambda_5)^2$ by the replacement $H\leftrightarrow A$.  

Care must be taken when applying the Higgs LET to diagrams containing gauge bosons. Indeed, and as discussed $e.g.$ in Refs.~\cite{Kniehl:1995tn,Pilaftsis:1997fe}, one must ensure that the quantity of which one is taking derivatives --- in our case the leading two-loop contributions to the photon self-energy --- is gauge independent. In our work, we choose to achieve this with two separate methods: on the one hand by employing the pinch technique~\cite{Cornwall:1981zr,Papavassiliou:1989zd,Degrassi:1992ue,Binosi:2002ez, Binosi:2002bs,Binosi:2009qm}, and on the other hand by using the background-field method (BFM)~\cite{DeWitt:1967ub,tHooft:1975uxh,Kluberg-Stern:1974nmx,Kluberg-Stern:1975ebk,Abbott:1980hw,Boulware:1980av,Hart:1983lbv,Denner:1994xt}. We confirmed that these two calculations produced exactly the same results for gauge-independent two-loop contributions to the photon self-energy, which we interpret as a strong cross-check of our expressions.  


Our calculation itself is divided into three main steps, described below. 
We have begun by generating genuine two-loop diagrammatic contributions to the photon self-energy using \texttt{FeynArts}~\cite{Kublbeck:1990xc,Hahn:2000kx}, using two different model files: one generated with \texttt{SARAH}~\cite{Staub:2009bi,Staub:2010jh,Staub:2012pb,Staub:2013tta} for the Feynmann gauge, and a second one devised for the BFM (adapting the results of Ref.~\cite{Denner:1994xt} to the IDM). The corresponding amplitudes were then computed and simplified with \texttt{FeynCalc}~\cite{Mertig:1990an,Shtabovenko:2016sxi,Shtabovenko:2020gxv}, and the reduction to master integrals was performed with \texttt{Tarcer}~\cite{Mertig:1998vk}. At intermediate stages of the computation, we retained the leading dependence on the light masses --- i.e.\ $m_h$, $m_G(=m_Z)$, $m_{G^\pm}(=m_W)$ --- as these serve to regulate infrared (IR) divergences in individual diagrams --- we will return to the cancellation of these IR divergences below. Once the contributions were expressed in terms of master integrals, we employed known expressions for the two-loop integrals (see $e.g.$ Refs.~\cite{Martin:2001vx,Martin:2003qz}). We also used expansions of the loop integrals around $p^{2}=0$, both with results available from the literature~\cite{Martin:2003qz,Braathen:2016cqe,Bahl:2021rts}, and from our own derivations using differential-equation based techniques (see for instance Refs.~\cite{Martin:2003qz,Bahl:2021rts}). All new expansions were verified numerically with the public tool \texttt{TSIL}~\cite{Martin:2005qm}.
In the Feynman gauge calculations, we include tadpole shifts so that tadpole contributions are canceled~\cite{Denner:1991kt} already at the level of bare perturbation theory.
This leads two-loop diagrams with tadpole shift in the NG-boson propagators.
In addition, we have included pinch terms obtained by the intrinsic pinch technique~\cite{Binosi:2002ez, Binosi:2002bs}.
On the other hand, on the BFM method, we have calculated the 1PI self-energy of classical photon field, which is manifestly gauge invariant. 
We have obtained the same expressions in both methods. 
In addition, our results are consistent with those in the two-Higgs doublet model calculated in the alignment limit and using the unitary gauge~\cite{Degrassi:2023eii}.

After obtaining closed-form expressions for the leading two-loop, unrenormalized, corrections to the photon self-energy --- expanded in terms of the ultraviolet (UV) regulator --- we inserted the full field dependence of masses and couplings, and took a derivative of these expressions with respect to the Higgs field $h$ in order to obtain the corresponding contributions to the Higgs di-photon decay amplitude. These results were then complemented by subloop renormalization contributions as well as external Higgs-leg corrections that also contribute at the leading two-loop level. The physical masses of the BSM scalars --- i.e.\ $m_H$, $m_A$, and $m_{H^\pm}$ --- 
as well as the EW VEV $v$ have been renormalized in an on-shell (OS) scheme (we note that for the renormalization of the VEV, we follow the prescription of Ref.~\cite{Kanemura:2015mxa}). For the BSM mass scale $\mu_2^2$, we have employed two different choices of renormalization schemes, a process-dependent on-shell (PDOS) scheme (following Refs.~\cite{Abe:2015rja,Banerjee:2021oxc}) as well as the decoupling-inspired (DI) scheme of Ref.~\cite{Braathen:2019pxr,Braathen:2019zoh}. Both options are discussed below together with the obtained analytical results.  

We have, in turn, performed a number of checks of our results, at different stages of the computation. First, we have verified that our results for the two-loop corrections to the photon self-energy obey the Ward-Takahashi identity of QED, which we did by confirming that $p^2\Pi_{\gamma\gamma}(p^2)\xrightarrow{p^2=0}0$, separately for each contribution. In the BFM, the QED Ward identity actually held independently for each diagram, while in the pinch-technique calculation the relation was fulfilled when combining all diagrams of a given class. As mentioned earlier, we also verified for each of the two-loop BSM contributions to the photon self-energy that the results obtained with the pinch technique and with the BFM were identical. Next, at the level of the decay amplitudes, we have verified that all UV divergences cancel between genuine two-loop terms and the one-loop subloop renormalization contributions. Moreover, we have confirmed that the renormalization scale dependence of individual diagrams cancels in the total results. 

Returning now to our choice of renormalization scheme for $\mu_2$, as a first option, we have implemented a process-dependent OS (PDOS) scheme, inspired by Refs.~\cite{Abe:2015rja,Banerjee:2021oxc}, in which the counterterm for $\mu_2^2$ is determined by requiring that the renormalized three-point function $\hat\Gamma_{hHH}$, evaluated with all three legs on shell (neglecting still $m_h$), should be equal to the tree-level result, $i.e.$, 
\begin{align}
\label{eq:PDOSscheme_def}
    \hat{\Gamma}_{hHH}(m_H^2,m_H^2,0)\overset{!}{=}\Gamma^\text{tree}_{hHH}\,.
\end{align}
In terms of the counterterm for $\mu_2^2$ --- which we denote in this scheme $(\delta^{(1)}\mu_2^2)^\text{PDOS}$ --- this implies that
\begin{align}
\label{eq:delmu2sq_PDOS}
    (\delta^{(1)}\mu_2^2)^\text{PDOS}=&\ \delta m_H^2\nn\\
   -\frac{1}{2}\Gamma_{hHH}^\text{tree} &v\left(-\frac{\delta v}{v}-\frac{1}{2}\Delta r+\frac{1}{2}\delta Z_h+\delta Z_H\right)\nn\\
   -\frac{1}{2}\Gamma_{hHH}^\text{1PI} &(m_H^2,m_H^2,0) v\,,
\end{align}
where $\Gamma_{hHH}^\text{1PI}$ denotes the one-particle-irreducible (1PI) one-loop corrections to the $\Gamma_{hHH}$ three-point function, $\delta Z_h$ and $\delta Z_H$ are the wave-function renormalization constants for $h$ and $H$ respectively, and $\delta v$ is the VEV counterterm and $\Delta r$ is the quantity that enters the relation between the OS-renormalized VEV and $G_F$~\cite{Kanemura:2019kjg,Sirlin:1980nh} (expressions for the one-loop IDM-specific contributions to the later two quantities are provided below). This PDOS scheme is also implemented in the program \texttt{H-COUP}~\cite{Kanemura:2017gbi,Kanemura:2019slf,Aiko:2023xui}, with which we performed our numerical evaluations. From \cref{eq:delmu2sq_PDOS}, we obtain
{\allowdisplaybreaks
\begin{widetext}
\begin{align}
\label{eq:dmu2_PDOS}
   (16\pi^2)(\delta^{(1)}\mu_2^2)^{\text{PDOS}}=&\ 3 \lambda_2 \mu_2^2\Delta_\text{UV}+\frac{(m_H^2 - \mu_2^2)\mhp{2}}{2 (m_A^2 -\mhp{2}) v^2}\bigg[m_A^2\log\frac{m_H^2}{m_A^2}+\frac{m_H^2 \mhp{2} + m_A^2 (\mhp{2}-2 m_H^2)}{m_H^2 - \mhp{2}}\log\frac{m_H^2}{\mhp{2}}\bigg]\nn\\
   &+\frac{\mu_2^2m_H^2}{v^2} \left(1 - \frac{m_A^2}{m_H^2}\right)^3 \text{Re}\bigg[\log\left(1 - \frac{m_H^2}{m_A^2}\right)\bigg]+ \frac{2 \mu_2^2m_H^2}{v^2}\left(1 - \frac{\mhp{2}}{m_H^2}\right)^3  \text{Re}\bigg[\log\left(1 - \frac{m_H^2}{\mhp{2}}\right)\bigg] \nn\\
   &+\frac{\mu_2^6}{6 v^2} \left(\frac{1}{m_A^2} - \frac{23}{m_H^2} + \frac{2}{\mhp{2}}\right)-\frac{\mu_2^4}{6 v^2}\left( \frac{m_H^2}{m_A^2} + \frac{2 m_H^2}{\mhp{2}}-63\right)+\frac{85 m_H^4}{12v^2}-\frac{23 m_H^2m_A^2}{12v^2} + \frac{m_A^4}{v^2}\nn\\
   &- \frac{23 m_H^2 \mhp{2}}{6v^2} + \frac{2 \mhp{4}}{ v^2}-\frac{\mu_2^2}{12 m_H^2 v^2} \Big[165 m_H^4 - 23 m_H^2 m_A^2 + 12 m_A^4- 46 m_H^2 \mhp{2} + 24 \mhp{4}\Big]\nn\\
   &+\frac{1}{2} \lambda_2 \big[3 m_H^2 + m_A^2 + 2 \mhp{2}\big]-\frac{1}{2}\lambda_2 \mu_2^2 \bigg[3 \log\frac{m_H^2}{Q^2} + \log\frac{m_A^2}{Q^2} + 2 \log\frac{\mhp{2}}{Q^2}\bigg]+\frac{7m_t^2}{2v^2} (m_H^2 - \mu_2^2)\,,
\end{align}
\end{widetext}
}\noindent where $\Delta_\text{UV}\equiv \frac{1}{\epsilon}-\gamma_E+\log(4\pi)$, with $\epsilon=(4-D)/2$ and $\gamma_E$ the Euler-Mascheroni constant. We note that we have defined $(\delta^{(1)}\mu_2^2)^{\text{PDOS}}$ so that it becomes a real quantity. As a second option, we also used the scheme proposed in Refs.~\cite{Braathen:2019pxr,Braathen:2019zoh}, which ensures the cancellation of the renormalization scale dependence and the proper decoupling of the BSM effects. Expressions for the $\mu_2^2$ counterterm, and for the contributions to the Higgs decay width to two photons in this second scheme are provided in \cref{app:2L_res_IDM_decoup}.

Lastly, a powerful check of the consistency of our results comes from confirming that all IR divergences cancel out. Indeed, individual topologies of diagrams contributing to the two-loop photon self-energy (and thus also to the effective Higgs-photon coupling) exhibit IR divergences caused by the light Higgs and NG bosons as well as gauge bosons, which are massless in our approximation. However, once all contributions of a given class of corrections (in terms of powers of scalar quartic couplings) have been summed, all IR divergences are canceled and the dependence on the light masses --- $m_h$, $m_G(=m_Z)$, $m_{G^\pm}(=m_W)$ --- that had been kept as regulators drops out entirely --- already at the level of unrenormalized self-energies. We note that this is different from what has been observed for scalar self-energies in relation with the Goldstone Boson Catastrophe (see e.g. Refs.~\cite{Elias-Miro:2014pca,Martin:2014bca,Kumar:2016ltb,Braathen:2016cqe,Braathen:2017izn,Goodsell:2019zfs}) where IR divergences are only cured by employing an OS scheme for the NG (and Higgs) boson masses, together the inclusion of external momentum.\footnote{Note that unlike the Higgs self-energies, for which the Goldstone Boson Catastrophe has been discussed in the literature, the photon self-energy is evaluated here at $p^2=0$, so that external momentum cannot serve to cure the IR divergences.}

To present our final analytical results, we decompose the di-photon decay width following
\begin{align}
\label{EQ:totaldecaywidth}
    \Gamma(h\to\gamma\gamma)=\ &\nn\\
    \frac{\sqrt{2}\alpha_\text{em}^2}{16\pi^3}G_Fm_h^3\bigg|&\mathcal{I}^{(1)}_t+\mathcal{I}^{(1)}_W+\mathcal{I}^{(1)}_{H^\pm}+\mathcal{I}^{(2)}_\text{SM, NLO QCD}\nn\\
    &+\mathcal{I}^{(3)}_\text{SM, NNLO QCD}+\mathcal{I}^{(2)}_\text{SM, NLO EW}\nn\\
    &+\mathcal{I}^{(2)}_{\ord{\lambda_3^2}}+\mathcal{I}^{(2)}_{\ord{(\lambda_4+\lambda_5)^2}}+\mathcal{I}^{(2)}_{\ord{(\lambda_4-\lambda_5)^2}}\nn\\
    &+\mathcal{I}^{(2)}_{\ord{\lambda_2}}+\mathcal{I}^{(2)}_\text{rem.}
    \bigg|^2\,,
\end{align}
where $\alpha_\text{em}=e^2/(4\pi)$ is the fine-structure constant and $G_F$ the Fermi constant.  
The first three terms correspond to the known one-loop corrections (see Ref.~\cite{Kanemura:2016sos}), while the next three terms, with subscripts ``SM'', are the higher-order QCD and SM-like EW corrections computed in Refs.~\cite{Djouadi:1990aj,Dawson:1992cy,Melnikov:1993tj,Djouadi:1993ji,Inoue:1994jq,Steinhauser:1996wy,Fleischer:2004vb,Djouadi:2005gi} and Refs.~\cite{Djouadi:1997rj,Fugel:2004ug,Aglietti:2004nj,Degrassi:2005mc,Passarino:2007fp,Actis:2008ts} respectively --- we note that for the latter we use the numerical value $\mathcal{I}^{(2)}_\text{SM, NLO EW}=\alpha_\text{em}/(16\pi s_W^2)(-24.1)$ from Ref.~\cite{Degrassi:2005mc}. The third and fourth lines contain the newly computed two-loop BSM contributions. 
$\mathcal{I}^{(2)}_{\ord{\lambda_3^2}},\ \mathcal{I}^{(2)}_{\ord{(\lambda_4\pm\lambda_5)^2}},\ \mathcal{I}^{(2)}_{\ord{\lambda_2}}$ are the genuine two-loop corrections respectively proportional to $\lambda_3^2$, $(\lambda_4\pm\lambda_5)^2$ or $\lambda_2$. Finally, $\mathcal{I}^{(2)}_\text{rem.}$ contains remaining pieces coming from the external-leg contributions $\mathcal{I}^{(2)}_\text{ext.-legs}$, terms arising from the renormalization of the EW VEV $\mathcal{I}^{(2)}_\text{VEV}$, and subloop renormalization contributions coming from parts of $(\delta^{(1)}\mu_2^2)^\text{PDOS}$ independent of $\lambda_2$. The expressions of the different new BSM contributions (together with the terms $\mathcal{I}^{(1)}_{H^\pm}$, $\mathcal{I}^{(1)}_t$, and $\mathcal{I}^{(1)}_W$, given here in the LET limit, to clarify the relative sign of one- and two-loop pieces) read
{\allowdisplaybreaks
\begin{widetext}
\begin{align}
\label{eq:2Lres_IDM}
 \mathcal{I}^{(1)}_{H^\pm}=&-\frac{1}{12}\left(1-\frac{\mu_2^2}{m_{H^\pm}^2}\right)\,,\qquad\qquad \mathcal{I}^{(1)}_t=-\frac{4}{9}\,,\qquad\qquad \mathcal{I}^{(1)}_W=\frac{7}{4}\,,\nn\\
 \mathcal{I}^{(2)}_{\ord{\lambda_3^2}}=&-\frac{1}{96\pi^2v^2}\left(1-\frac{\mu_2^2}{\mhp{2}}\right)^2(2\mhp{2}+\mu_2^2)\,,\nn\\
 \mathcal{I}^{(2)}_{\ord{(\lambda_4+\lambda_5)^2}}=& -\frac{1}{192\pi^2 v^2}\bigg\{\mhp{2} \left(1-\frac{m_H^2}{\mhp{2}}\right)^2+ \mu_2^2\left(\frac{m_H^4}{\mhp{4}} - \frac{3 m_H^2}{ \mhp{2}} +38\right)\nn\\
 &\hspace{1.75cm}+\mu_2^2 \bigg[-\frac{17 m_H^2 + 19\mhp{2}}{m_H^2 - \mhp{2}}\log\frac{m_H^2}{\mhp{2}}+ \left(\frac{m_H^2}{\mhp{2}} - 1\right)^3 \mathrm{Re}\left[\log\left(1 - \frac{\mhp{2}}{m_H^2}\right)\right]\bigg]\bigg\}\,,\nn\\
  \mathcal{I}^{(2)}_{\ord{(\lambda_4-\lambda_5)^2}}=&-\frac{1}{192\pi^2 v^2}\bigg\{\mhp{2} \left(1-\frac{m_A^2}{\mhp{2}}\right)^2+ \mu_2^2\left(\frac{m_A^4}{\mhp{4}} - \frac{3 m_A^2}{\mhp{2}} +38\right)\nn\\
  &\hspace{1.75cm}+\mu_2^2 \bigg[-\frac{17 m_A^2 + 19\mhp{2}}{m_A^2 - \mhp{2}}\log\frac{m_A^2}{\mhp{2}}+ \left(\frac{m_A^2}{\mhp{2}} - 1\right)^3 \mathrm{Re}\left[\log\left(1 - \frac{\mhp{2}}{m_A^2}\right)\right]\bigg]\bigg\}\,,\nn\\
  \mathcal{I}^{(2)}_{\ord{\lambda_2}}\bigg|^{ (\delta^{(1)}\mu_2^2)^\text{PDOS}}=&-\frac{\lambda_2}{192\pi^2\mhp{2}}\bigg(\mhp{2}-m_H^2 + \mu_2^2 \log\frac{m_H^2}{\mhp{2}}\bigg)\,\nn\\ 
  \mathcal{I}^{(2)}_\text{ext.-legs}+\mathcal{I}^{(2)}_\text{VEV}=&\ \frac{7m_t^2}{32\pi^2v^2}\mathcal{I}^{(1)}_{H^\pm}+\left[\mathcal{I}_t^{(1)}+\mathcal{I}^{(1)}_{H^\pm}+\mathcal{I}^{(1)}_W\right]\bigg(\frac{1}{2}\Sigma_{hh}^{(1)\ \prime}(0)|^\text{BSM}-\frac{\delta v}{v}\bigg|^\text{BSM}-\frac{1}{2}\Delta r|^\text{BSM}\bigg)\,.\nn\\
  \mathcal{I}^{(2)}_\text{rem.}\equiv&\ \mathcal{I}^{(2)}_\text{ext.-legs}+\mathcal{I}^{(2)}_\text{VEV}+\frac{\partial\mathcal{I}^{(1)}_{H^\pm}}{\partial \mu_2^{2}}\left((\delta^{(1)}\mu_2^2)^\text{PDOS}\Big|_{\lambda_2\to0}\right)\nn\\
  =&\ \frac{\mu_2^2m_H^2}{192\pi^2 v^2\mhp{2}}\left[\left(1-\frac{m_A^2}{m_H^2}\right)^3\Re\qty[\log\left(1-\frac{m_H^2}{m_A^2}\right)]+2\left(1-\frac{\mhp{2}}{m_H^2}\right)^3\Re\qty[\log\left(1-\frac{m_H^2}{\mhp{2}}\right)]\right]\nn\\
 &+\frac{3 m_H^2 + 44 \mhp{2}}{1152 \pi^2 v^2 (m_A^2 - \mhp{2})}\left[m_A^2\log\frac{m_H^2}{m_A^2}+\frac{m_H^2 \mhp{2} + m_A^2 (\mhp{2}-2 m_H^2)}{m_H^2 - \mhp{2}}\log\frac{m_H^2}{\mhp{2}}\right]\nn\\
 &+\frac{7m_t^2 (m_H^2 - \mhp{2})}{384 \pi^2 v^2\mhp{2}}+\frac{255 m_H^4 - 94 m_H^2 \mhp{2} + 160 \mhp{4} + 36 m_A^4 + m_A^2 (-69 m_H^2 + 44 \mhp{2})}{6912 \pi^2 v^2\mhp{2}}\nn\\
 &-\frac{\mu_2^2}{1728 \pi^2 v^2 m_H^2\mhp{2}} \Big[123 m_H^4 - 18 m_H^2m_A^2 + 9 m_A^4 - 212 m_H^2 \mhp{2} + 18 \mhp{4}\Big] \nn\\
 &-\frac{\mu_2^4}{3456\pi^2 v^2 m_H^2 m_A^2 \mhp{4}}\Big[ 6 m_H^4m_A^2 + m_H^2\mhp{2} (3 m_H^2-125 m_A^2) + 44 \mhp{4}(m_H^2 + m_A^2)\Big] \nn\\
 &-\frac{\mu_2^6}{48\pi^2 v^2 m_H^2 \mhp{2}} \,.
\end{align}
\end{widetext}
}
In the equations above, $\Sigma_{hh}^{(1)}(0)|^\text{BSM}$ and $\delta v/v|^\text{BSM}$ denote the BSM contributions respectively to the one-loop Higgs boson self-energy and to the VEV counter-term. The former can be found to read
\begin{align}
    \Sigma_{hh}^{(1)\ \prime}&(0)|^\text{BSM}=    -\frac{1}{48\pi^2v^2}\Bigg[m_H^2\left(1-\frac{\mu_2^2}{m_H^2}\right)^2\\
    &+m_A^2\left(1-\frac{\mu_2^2}{m_A^2}\right)^2+2\mhp{2}\left(1-\frac{\mu_2^2}{\mhp{2}}\right)^2\Bigg]\,,\nn
\end{align}
while the latter is
\begin{align}
\label{eq:VEV_CT_BSM}
    &\frac{\delta^{(1)}v^\text{OS}}{v^\text{OS}}\bigg|^\text{BSM}=\frac{1}{64\pi^2s_W^2v^2}\bigg\{-c_W^2\mathcal{F}(m_H^2,m_A^2)\nn\\
    &\quad+(1 - 2 s_W^2)\big[\mathcal{F}(\mhp{2},m_H^2)+\mathcal{F}(\mhp{2},m_A^2)\big]\bigg\}\,,
\end{align}
with 
\begin{align}
    \mathcal{F}(x,y)\equiv\frac{x^2-y^2+2xy\log\frac{y}{x}}{x-y}\,,
\end{align}
and with $c_W$ and $s_W$ the cosine and sine of the weak mixing angle. Finally, we use the $G_F$ scheme for the EW input parameters, and therefore we include an additional piece in the results arising from $\Delta r|^\text{BSM}$,  $i.e.$ the leading one-loop BSM contributions to the quantity $\Delta r$ that appears in the relation between the OS-renormalized VEV and $G_F$ --- see e.g. Ref.~\cite{Sirlin:1980nh,Kanemura:2019kjg}. The expression of $\Delta r|^\text{BSM}$ reads
\begin{align}
    \Delta r|^\text{BSM}=\frac{c_W^2}{32\pi^2s_W^2v^2}&\bigg\{\mathcal{F}(m_H^2, m_A^2) - \mathcal{F}(m_H^2, m_{H^\pm}^2) \nn\\
    &- \mathcal{F}(m_A^2, m_{H^\pm}^2)\bigg\}\,.
\end{align}
We note, however, that the scenarios investigated in the following section, for which we set $m_{H^\pm}=m_A$, the leading BSM contributions  $\Delta r|^\text{BSM}$ vanish. 

\section{Numerical results}
\label{sec:numres}
We present in this section numerical investigations of our new results and their phenomenological impact. We begin by defining two benchmark scenarios, fulfilling all the theoretical and experimental constraints discussed in \cref{sec:IDM} and inspired by DM phenomenology --- following also Ref.~\cite{Kanemura:2016sos}. We set for both scenarios $m_{H^\pm}=m_A$, so that the custodial symmetry is restored in the scalar sector --- thereby ensuring that EWPOs are in good agreement with their experimentally measured results. As can be seen from \cref{eq:0L_masses}, this choice also implies that $\lambda_4-\lambda_5=0$, so that corrections of $\ord{(\lambda_4-\lambda_5)^2}$ will not appear --- however, we emphasize once again that the form and behavior of these effects are analogous to that of the $\ord{(\lambda_4+\lambda_5)^2}$ pieces, which will be present in our investigations. We furthermore fix the BSM mass parameter $\mu_2$ for each scenario from the requirement that the DM relic density (computed with \texttt{micrOMEGAs\_6.1.15}~\cite{Alguero:2023zol}) should not exceed the value measured by PLANCK~\cite{Planck:2018vyg}, while simultaneously evading direct detection limits. Throughout this section, the values given for $\mu_2$ will be understood as those in the PDOS scheme. 

A first scenario, which we will refer to as the \textit{Higgs resonance scenario}, is defined by
\begin{align}
\label{eq:HRsc}
    &\mu_2^2=3581\text{ GeV}^2\,,\quad m_H=60\text{ GeV}\,,\nn\\
    &100\text{ GeV}\leq m_{H^\pm}(=m_A)\leq620\text{ GeV}\,.
\end{align}
The lower bound on $m_{H^\pm}(=m_A)$ comes from direct searches at LEP~\cite{Lundstrom:2008ai}. To ensure perturbativity, we remove the parameter points with $\abs{\lambda_{i}}>4\pi$ even if they are allowed by tree-level perturbative unitarity\footnote{Considering only tree-level perturbative unitarity, we could take $m_{H^\pm}\simeq 700\text{ GeV}$, which corresponds to $\lambda_3=16.0$ and $\lambda_4=\lambda_5=-8.02$.}. These mass ranges correspond to Lagrangian quartic couplings ranging from $\lambda_3=0.212,\ \lambda_4=\lambda_5=-0.106$ (for $m_{H^\pm}=100\text{ GeV}$) to $\lambda_3=12.6,\ \lambda_4=\lambda_5=-6.28$ (for $m_{H^\pm}=600\text{ GeV}$). The scalar DM candidate $H$ has a mass of approximately $m_h/2$, which leads to an enhancement of the DM relic density via Higgs resonance (see $e.g.$ Ref.~\cite{Belyaev:2016lok}) --- hence the name of this scenario. Due to the proximity of $\mu_2$ and $m_H$, the branching ratio for the invisible decay $h\to HH$ is about $0.01\%$ in this scenario --- well below the current bounds (see $e.g.$ Refs.~\cite{ATLAS:2022vkf,CMS:2022dwd,ATLAS:2023tkt,CMS:2023sdw,Biekotter:2022ckj}) as well as expected limits at $e^+e^-$ Higgs factories like the ILC~\cite{Fujii:2017vwa} or the FCC-ee~\cite{deBlas:2019rxi}. We have also verified that the spin-independent direct detection cross-section~\cite{Alguero:2023zol} is $\sigma_{SI}=9.4\cdot10^{-49} \text{ cm}^2$ in this scenario (we note that, at leading order, this cross-section only depends on $m_H$ and $\mu_2$), so that it fulfills the upper bound from the recent LZ results~\cite{LZ:2024} --- which corresponds to $\sigma_{SI}\lesssim 2\cdot10^{-48}\text{ cm}^2$ for $m_H=60\text{ GeV}$. The value of $\Omega_{\rm DM}h^{2}$ is about 0.12 and is approximately independent of $m_{H^{\pm}}$ and $m_{A}$ for most of the considered mass range. We note that since the evaluation of $\Omega_{\rm DM}h^{2}$ is only performed at tree level, we restrict ourselves to a level of accuracy of two digits. 

We also consider a second scenario where all BSM scalars are heavy (we refer to this as the \textit{heavy Higgs scenario}). This case is less directly related to DM phenomenology, as DM (for which the candidate state is still taken to be $H$) is typically underproduced. We set
\begin{align}
\label{eq:HHsc}
    &\mu_2=499.9\text{ GeV}\,,\quad m_H=500\text{ GeV}\,,\nn\\
    &500\text{ GeV}\leq m_{H^\pm}(=m_A)\leq790\text{ GeV}\,.
\end{align}
The lower bound on $m_{H^\pm}(=m_A)$ is this time motivated by vacuum stability --- see \cref{eq:vac_stab_2} --- which implies that all BSM scalar masses should be larger than $|\mu_2|$. The upper bound is once again obtained from requiring $\abs{\lambda_{i}}\leq 4\pi$. (in the most favorable case of $\lambda_2\simeq0$). In terms of Lagrangian quartic couplings, this corresponds to values ranging from $\lambda_3=3.30\times10^{-3},\ \lambda_4=\lambda_5=0$ (for $m_{H^\pm}=500\text{ GeV}$) to $\lambda_3=12.3$ and $\lambda_4=\lambda_5=-6.17$ (for $m_{H^\pm}=790\text{GeV}$). For both scenarios, $\lambda_2$ is kept as a free parameter, and we choose several example values $\lambda_2=0.1,\ 1,\ 5$ and adapt the upper limit of the range of $m_{H^\pm}(=m_A)$ allowed under perturbativity accordingly (the bound on the masses is lower for higher $\lambda_2$). We also note that, in this scenario, decays of the Higgs boson into pairs of inert scalars are kinematically forbidden, so that there is no constraint from invisible Higgs decays. Once again, we have verified for this scenario the upper limit on the spin-independent direct detection cross-section, which we find to be $\sigma_{SI}=3.9\cdot 10^{-49}\text{ cm}^2$, below the bound from LZ~\cite{LZ:2024} of $\sigma_{SI}\lesssim 1\cdot 10^{-47}\text{ cm}^2$ for 
$m_H=500\text{ GeV}$.   

In the following, we present our results for the Higgs decay to two photons in terms of the ratio
\begin{align}
    \brr\equiv \frac{\text{BR}(h\to\gamma\gamma)_\text{IDM}}{\text{BR}(h\to\gamma\gamma)_\text{SM}}\,,
\end{align}
$i.e.$ the ratio of the branching ratio of the di-photon decay of the Higgs boson computed in the IDM over that in the SM. The branching ratios are computed using the program \texttt{H-COUP}~\cite{Kanemura:2017gbi,Kanemura:2019slf,Aiko:2023xui}, supplemented by the NLO (two-loop) corrections to the di-photon decay width given in the previous section. We note that the total Higgs boson width, to which the partial decay widths $\Gamma(h\to\gamma\gamma)$ are normalized, is always calculated at NLO --- this choice allows to disentangle the effect of the newly computed NLO corrections to the di-photon decay from known one-loop corrections to other Higgs decay channels. This quantity can also be compared directly to current experimental measurements~\cite{CMS:2021kom,ATLAS:2022tnm,ATLAS:2022vkf,CMS:2022dwd} or future prospects~\cite{Cepeda:2019klc} --- $e.g.$ it corresponds to the quantity $B^{\gamma\gamma}$ in Ref.~\cite{Cepeda:2019klc}. In order to better understand and illustrate the size of the two-loop corrections to the di-photon partial decay width, we will also investigate in the following a second quantity, \gammar, which we define to be
\begin{align}
    \gammar\equiv \frac{\Gamma(h\to\gamma\gamma)_\text{IDM}}{\Gamma(h\to\gamma\gamma)_\text{SM}}\,.
\end{align}
In this ratio, the SM pieces approximately cancel out between the numerator and denominator, so that we can directly assess the magnitude of BSM effects. 

\begin{figure*}
\includegraphics[width=0.48\textwidth]{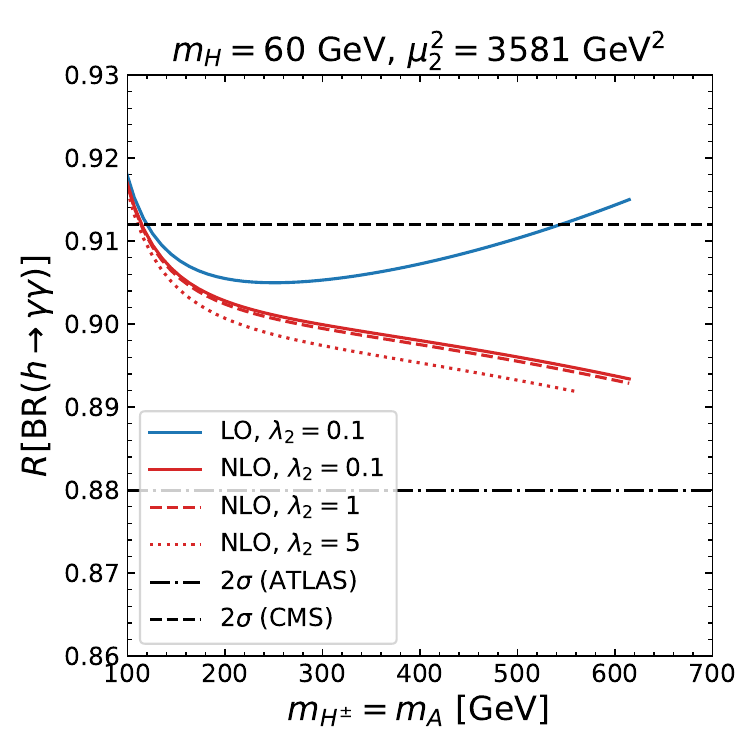}\hspace{5mm}
\includegraphics[width=0.48\textwidth]{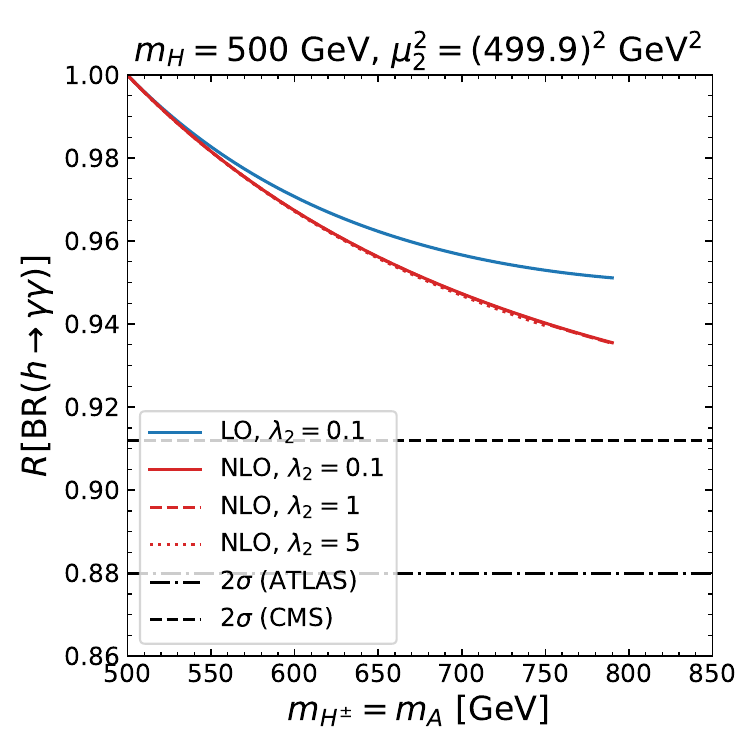}
	\caption{Prediction for \brr~as a function of $m_{H^{\pm}}=m_A$. \textit{Left}: Higgs resonance scenario, defined in \cref{eq:HRsc} \textit{Right}: Heavy Higgs scenario, defined in \cref{eq:HHsc}. }
	\label{fig:Rgamgam_mch}
\end{figure*}

\begin{figure*}
\includegraphics[width=0.48\textwidth]{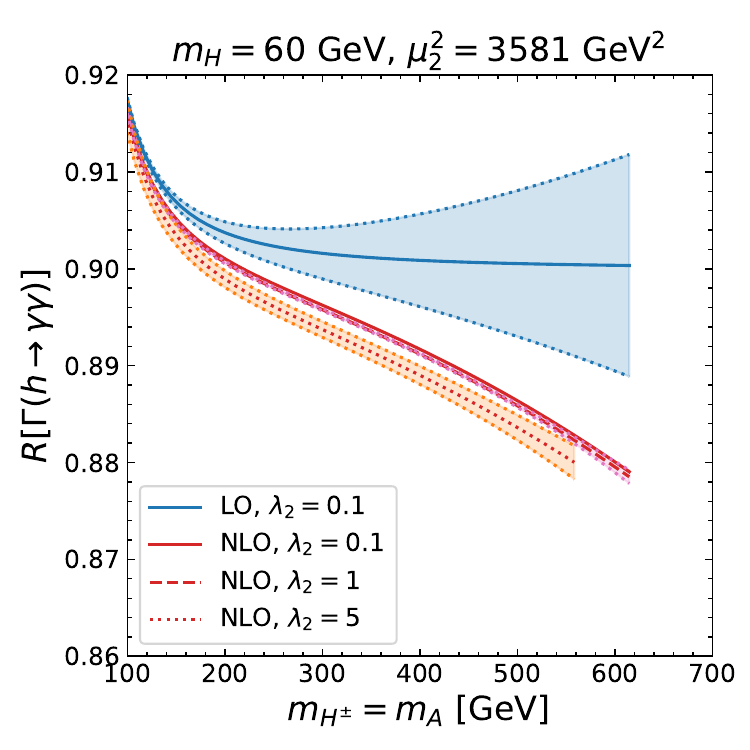}\hspace{5mm}
\includegraphics[width=0.48\textwidth]{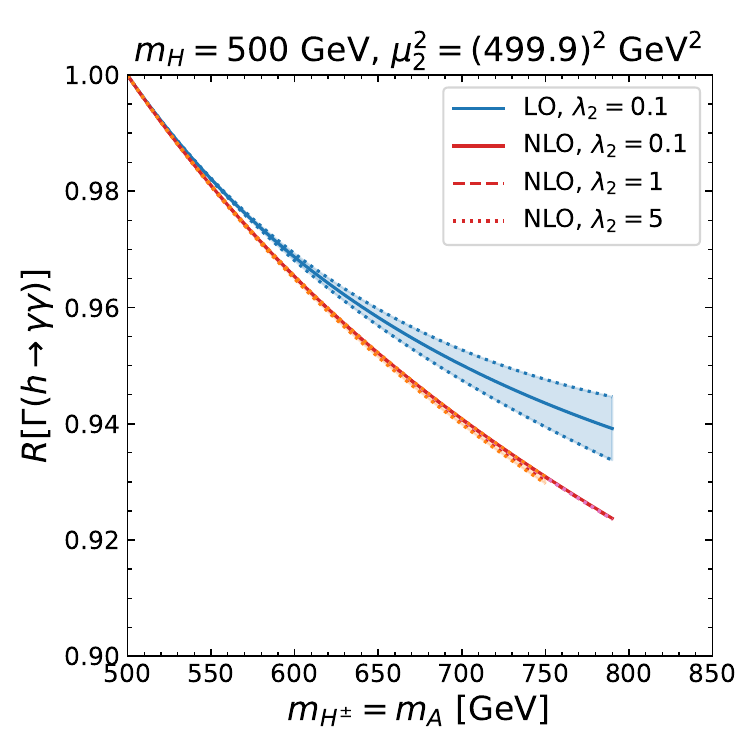}
	\caption{Prediction for \gammar~as a function of $m_{H^{\pm}}=m_A$. \textit{Left}: Higgs resonance scenario, defined in \cref{eq:HRsc} \textit{Right}: Heavy Higgs scenario, defined in \cref{eq:HHsc}. The colored bands represent uncertainty estimates at one- and two-loop levels and are explained in the text.}
	\label{fig:dkappagamgam_mch}
\end{figure*}

In \cref{fig:Rgamgam_mch,fig:dkappagamgam_mch}, we show respectively results for \brr~and \gammar~as a function of $m_{H^\pm}(=m_A)$ for the Higgs resonance (left) and the heavy Higgs (right) scenarios. The leading order ($i.e.$ one-loop) results are given by the blue curves, while the red curves correspond to the NLO ($i.e.$ two-loop) results for different values of the inert quartic coupling $\lambda_2$. The black lines in \cref{fig:Rgamgam_mch} indicate the expected 95\% CL bounds on the ratio \brr~at HL-LHC~\cite{Cepeda:2019klc}: the dot-dashed line corresponds to the more conservative limit from ATLAS, while the dashed line is the stronger limit from CMS (which has a better detector for photons). We note that the corresponding current LHC bounds~\cite{CMS:2021kom,ATLAS:2022vkf,CMS:2022dwd} correspond to $\brr=0.8$ and are outside the plots. Turning first to the LO results, we can observe that the BSM deviation in \gammar~(\cref{fig:dkappagamgam_mch}) approaches a plateau as $m_{H^\pm}$ increases as can be seen from $\mathcal{I}^{(1)}_{H^\pm}$ in Eq.~\eqref{eq:2Lres_IDM} (for the heavy Higgs scenario, this plateau is however not reached due to the limit from perturbativity). This behavior can be explained by the compensation between the charged-Higgs mass dependence in the coupling $\lambda_3\propto (m_{H^\pm}^2-\mu_2^2)$ and in the loop function for the charged Higgs loop at LO --- see for instance equations (A.34)-(A.35) in Ref.~\cite{Kanemura:2016sos}. In the left plot of \cref{fig:Rgamgam_mch}, one can observe an increase in {\brr} for $m_{H^\pm}\gtrsim 250\text{ GeV}$. This is because the ratio $(\Gamma^\text{tot}_h)_\text{IDM}/(\Gamma^\text{tot}_h)_\text{SM}$ decreases as $m_{H^\pm}$ increases~\cite{Kanemura:2019kjg}. At two loops, the behavior of the two-loop corrections to $\Gamma(h\to\gamma\gamma)$ is drastically modified, and in both scenarios they continue growing with $m_{H^\pm}$, increasing the deviation from the SM. On the one hand, the parametric dependence of the two-loop corrections --- $c.f.$ \cref{eq:2Lres_IDM} --- is different, so that these still grow for fixed $\mu_2$ and increasing $m_{H^\pm}$. On the other hand, several new types of contributions arise at two loops that are not corrections of the LO BSM effects --- specifically the $\ord{\lambda_3^2}$ terms can be understood as a correction of the LO $\ord{\lambda_3}$ charged-Higgs loop, while the $\ord{(\lambda_4+\lambda_5)^2}$ and $\ord{\lambda_2}$ terms as well as the external-leg corrections correspond to new classes of effects only entering $\Gamma(h\to\gamma\gamma)$ from two loops. Importantly, we find that the inclusion of two-loop corrections in the di-photon decay width increases the size of the BSM deviation. To be concrete, we can consider the situation in the Higgs resonance scenario for $m_{H^\pm}=500$ GeV: we find that the BSM deviation in {\brr} of about $-9\%$ at one loop increases to about $-10.4\%$ for low values of $\lambda_2$ ($=0.1$ or $1$) or almost $-10.8\%$ for large $\lambda_2=5$. In the heavy Higgs scenario and for $m_{H^\pm}=700$ GeV, it is about $-4.5\%$ at one loop and grows to $-5\%$ at two loops (for all the values of $\lambda_2$ considered here). For the Higgs resonance scenario, the BSM deviation \brr, computed at the two-loop level, is larger than the expected CMS sensitivity at HL-LHC for most of the mass range (for $\mhp{}\gtrsim 120$ GeV), while with the one-loop calculation, only the interval $120\text{ GeV}\lesssim \mhp{}\lesssim 550\text{ GeV}$ would be above the CMS sensitivity. On the other hand, no point in the allowed mass range of the Higgs resonance scenario would produce a large enough deviation to exceed the expect ATLAS sensitivity at HL-LHC. In the heavy Higgs scenario, neither ATLAS nor CMS would have sensitivity to BSM deviations in \brr, throughout the allowed mass range. The picture would, however, be modified with a future $e^+e^-$ Higgs factory, like ILC or FCC-ee, at which the improvement in the determination of the total Higgs boson width would result in a better experimental precision on \brr~--- a precision on this branching ratio of about $2.6\%$~\cite{Fujii:2017vwa} could be attained at ILC-500, or of about $3\%$~\cite{FCC:2018byv,FCCICHEP:2024} at FCC-ee (while further improvements could also be achieved at a high-energy $e^+e^-$ collider or at a muon collider, see e.g.\ Ref.~\cite{Forslund:2022xjq}). 
Such a level of precision would allow probing BSM deviations in \brr~in the high mass range of the heavy Higgs scenario --- but a reliable interpretation of a possible detection of a BSM deviation in \brr~can only be achieved in this scenario provided that two-loop corrections to $\Gamma(h\to\gamma\gamma)$ are included. In the Higgs resonance scenario, the entire mass range from \cref{eq:HRsc} could be excluded at the $2\sigma$ level. Moreover, the difference between the one- and two-loop curves is even of the order, or larger, than the expected accuracy at $e^+e^-$ colliders. Therefore, in both scenarios the inclusion of two-loop corrections is necessary to properly interpret the observation or non-observation of BSM deviations in \brr~in terms of the parameter space of the IDM. 

Reliable interpretations of theoretical predictions additionally require corresponding uncertainty estimates. Therefore, we employ renormalization scheme conversions of the BSM mass parameter $\mu_2$ between the PDOS and DI schemes (described in \cref{sec:calc}) in order to estimate the size of unknown higher-order contributions, which are not included in our calculations. The results we obtain are shown as bands in \cref{fig:dkappagamgam_mch}, symmetrized around the predictions for $R[\Gamma(h\to\gamma\gamma)]$ in the PDOS scheme. Specifically, the light blue band corresponds to the scheme conversion of $\mu_2$ in the one-loop prediction (using $\lambda_2=0.1$ in the conversion), while the magenta and orange bands show the impact of the conversion\footnote{We note that conversions of $\mu_2$ in the one-loop result for $\Gamma(h\to\gamma\gamma)$ do not generate terms of all the possible terms present at two loops. Thus, to ensure meaningful comparisons of the one- and two-loop uncertainty bands, we have restricted the conversion at two loops to the types of contributions that are generated by one-loop conversions --- i.e.\ the term $\mathcal{I}^{(2)}_{\mathcal{O}(\lambda_2)}$ and the purely BSM corrections in $\mathcal{I}^{(2)}_\text{rem.}$. } at two loops, for $\lambda_2=1$ (magenta) and $\lambda_2=5$ (orange) --- and we use the expressions in the DI scheme, in \cref{eq:I2_DIscheme}, for the two-loop predictions of $R[\Gamma(h\to\gamma\gamma)]$ after the scheme conversion. It should be emphasized that the bands obtained with the conversion of $\mu_2$ alone should not be taken as complete theory uncertainties, but as \emph{estimates} of the size of missing higher-order effects. As expected, the size of the unknown higher-order contributions is reduced when going from the one- to the two-loop level; for instance, considering the Higgs resonance scenario (left panel) and $m_{H^\pm}=500\text{ GeV}$, the size of the band, in relative size compared to the PDOS prediction, is reduced from $\pm 8.4\%$ at one loop to $\pm0.16\%$ ($\pm1.1\%$) at two loops for $\lambda_2=1$ ($\lambda_2=5$). The difference between the two values at two loops can simply be understood as being due to the impact of increasing values of $\lambda_2$ on the magnitude of the loop corrections. These findings constitute a positive indication of the convergence of the perturbative expansion performed in our loop calculations, as well as of the high level of accuracy with which our results can be compared with experimental results.

\begin{figure*}
    \centering
    \includegraphics[width=0.48\textwidth]{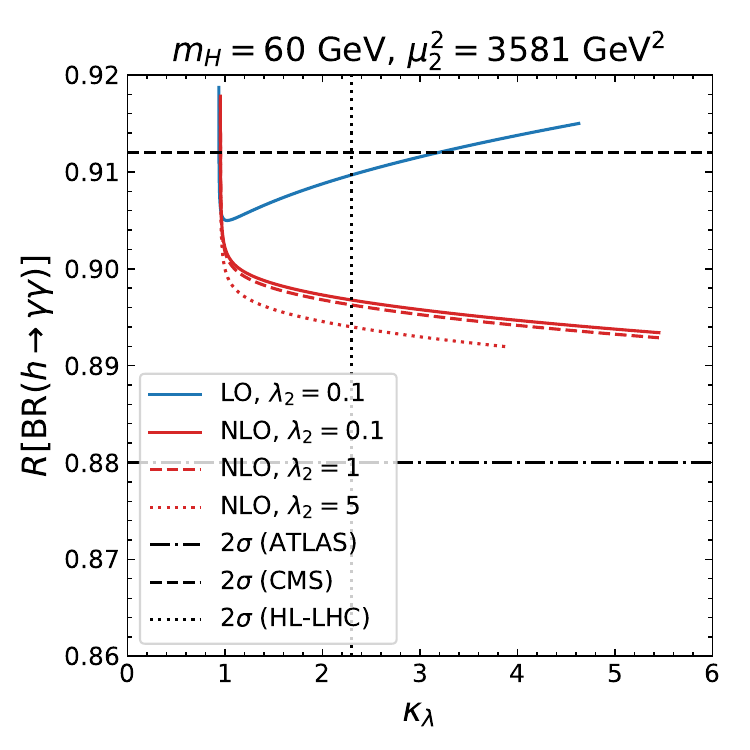}\hspace{5mm}
    \includegraphics[width=0.48\textwidth]{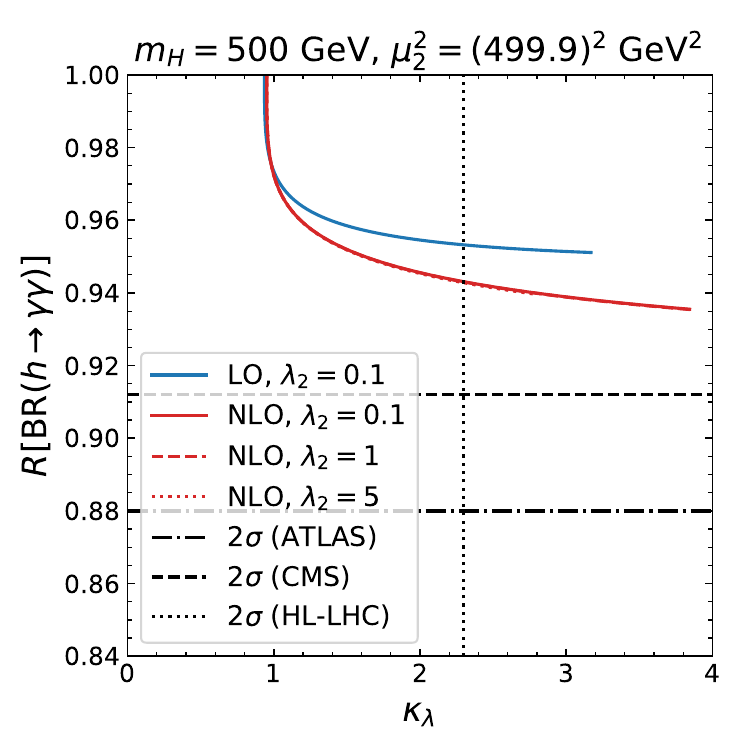}
    \caption{
    Correlation between $\kappa_\lambda$ and \brr~at one loop (blue) and two loops (red). \textit{Left}: DM-inspired Higgs resonance scenario. \textit{Right}: Heavy Higgs scenario. }
    \label{fig:correlationplot}
\end{figure*}

Correlations between BSM deviations in various couplings of the 125-GeV Higgs boson can play a crucial role in identifying, or ``fingerprinting,'' the nature of the underlying BSM physics at the origin of the deviations --- this was for instance pointed out in Ref.~\cite{Kanemura:2015mxa} and studied at NLO in Refs.~\cite{Kanemura:2019kjg, Aiko:2021nkb}. However, in this context, it is also important to assess how higher-order corrections to the different couplings can modify the correlations found at LO. Therefore, we illustrate in \cref{fig:correlationplot} the interplay between \brr~and $\kappa_\lambda$ and how it is modified when going from LO to NLO in the calculation of BSM corrections to these two quantities.\footnote{We note that BSM corrections to the trilinear Higgs coupling only occur at the loop level. As was pointed out in e.g.\ Refs.~\cite{Kanemura:2002vm,Kanemura:2004mg,Braathen:2019zoh,Bahl:2022jnx}, these effects are not a correction of the tree-level prediction for this coupling, but rather a new class of contributions entering from the one-loop level. We therefore refer to the one- and two-loop BSM corrections to $\kappa_\lambda$ as respectively LO and NLO effects. }$\kappa_\lambda$ denotes here the coupling modifier for the trilinear Higgs coupling, defined\footnote{This definition corresponds to the effective trilinear Higgs coupling that is constrained by the ATLAS and CMS collaborations via double- and single-Higgs production~\cite{ATLAS:2022kbf,ATLAS:2022vkf,CMS:2022dwd}. } as $\kappa_\lambda\equiv \lambda_{hhh}^\text{IDM}/(\lambda_{hhh}^\text{SM})^{(0)}$, with $\lambda_{hhh}^\text{IDM}$ the trilinear Higgs coupling calculated at one and two loops in the IDM and with $(\lambda_{hhh}^\text{SM})^{(0)}$ the tree-level prediction for this coupling in the SM. As in the previous figure, we consider both benchmark scenarios (Higgs resonance on the left and heavy Higgs on the right), and the blue and red curves correspond to one- and two-loop results respectively. Each of these lines is obtained by varying $m_{H^\pm}=m_A$ in the ranges given in \cref{eq:HRsc,eq:HHsc}. For the values of $\kappa_\lambda$ shown in \cref{fig:correlationplot}, we perform a full one-loop calculation of $\kappa_\lambda$ with the public tool \texttt{anyH3}~\cite{Bahl:2023eau}, which we complement at the two-loop level by employing results from Refs.~\cite{Braathen:2019pxr,Braathen:2019zoh}. These results were derived for the Higgs resonance scenario ($i.e.$ they include only the dependence on $m_A$, $m_{H^\pm}$ and $\lambda_2$), and we have moreover extended them for use in the heavy Higgs scenario by including also the full dependence on $m_H$ and $\mu_2$ in $\lambda_{hhh}$ --- we provide the new expressions in \cref{app:2L_hhh}. Black lines indicate the expected limits at the HL-LHC~\cite{Cepeda:2019klc} for \brr~(the dot-dashed line corresponds to the ATLAS limit, and the dashed line to the CMS one, similarly to Fig~\ref{fig:Rgamgam_mch}) and $\kappa_\lambda$ (dotted line, obtained from a combination of ATLAS and CMS expected bounds). The ranges of values for $\kappa_\lambda$ shown on the horizontal axis in both plots of \cref{fig:correlationplot} are within the interval $-1.2<\kappa_\lambda<7.2$, which are the current bounds on this effective coupling~\cite{ATLAS:2022tnm,ATLAS:2022vkf,CMS:2022dwd,ATLAS:2022kbf,ATLAS:2024ish} (we note that we quote here the recent upper bound from Ref.~\cite{ATLAS:2024ish}, which takes into account theoretical uncertainties in the di-Higgs production cross-section in a more reliable manner than earlier results). The reason why some of the curves in \cref{fig:correlationplot} stop before the end of the range of values of $\kappa_\lambda$ shown on the horizontal axis is again the criterion of perturbativity, which sets an upper limit on the allowed range of masses $m_{H^\pm}(=m_A)$ and thus in turn on the possible values of $\kappa_\lambda$ at one and two loops. 

For the lower end of the mass ranges in both scenarios ($i.e.$ in the upper left corner of the plots), the BSM effects in \brr~grow faster, in relative values, than those in $\kappa_\lambda$. However, the size of the deviations in the latter are of course much larger, with effects of several hundred percent being possible --- as observed also in $e.g.$ Refs.~\cite{Kanemura:2004mg,Braathen:2019pxr,Braathen:2019zoh,Bahl:2022jnx}. In the heavy Higgs scenario (right plot), the two-loop corrections to \brr~are not significant enough to modify the correlation with $\kappa_\lambda$ drastically. Moreover, as seen already in \cref{fig:Rgamgam_mch}, even at the HL-LHC, it seems difficult to access BSM deviations in \brr~for this scenario, and $\kappa_\lambda$ appears to offer better prospects of probing the IDM with heavy inert scalars --- in particular, at the HL-LHC the mass range above about 730 GeV could be probed. On the other hand, the situation is once again more interesting for the Higgs resonance scenario (left plot). The two-loop corrections to \brr~are here sizeable, and the shape of the \brr~vs $\kappa_\lambda$ curves are noticeably modified by the inclusion of two-loop corrections. Therefore, it is clearly necessary to include all known higher-order BSM corrections to both quantities to reliably investigate their interplay. Furthermore, while at present one does not expect to see any deviation in \brr~and in $\kappa_\lambda$ only the upper end of the mass range is accessible, almost the entire mass range in this scenario --- $i.e.$ $m_{H^\pm}\gtrsim 120$ GeV --- can be probed at the HL-LHC (considering the expected bound from CMS), and \brr~appears to be more sensitive than $\kappa_\lambda$ for lower masses. Finally, we note that combined measurements of \brr~and $\kappa_\lambda$ offer the interesting possibility --- especially in the Higgs resonance scenario --- of obtaining some information about the inert quartic coupling $\lambda_2$ that is otherwise very difficult to probe.

\section{Discussion and conclusion}
\label{sec:ccl}
We have performed in this paper the first calculation of leading two-loop (NLO) BSM corrections to the decay of a Higgs boson into two photons in the IDM. In line with our goal of assessing the size of the dominant two-loop effects, we have considered only contributions involving inert scalars and gauge bosons, and we have neglected the dependence on light scalar masses. These assumptions allowed us to use a Higgs LET to obtain compact analytic expressions for the leading two-loop corrections to $\Gamma(h\to\gamma\gamma)$. To ensure that we obtained gauge-independent results for the two-loop corrections to the photon self-energy, which we employ in the Higgs LET, we performed two separate calculations with the background-field method as well as the pinch technique --- finding full agreement between the two results. Additionally, we employed the on-shell renormalization scheme for the BSM scalar masses and the EW vacuum expectation value. For the BSM mass parameter $\mu_2$, we have employed two different choices of renormalization schemes, a process-dependent OS scheme following Refs.~\cite{Abe:2015rja,Banerjee:2021oxc} and that is implemented in \texttt{H-COUP}~\cite{Kanemura:2017gbi,Kanemura:2019slf,Aiko:2023xui}, as well as the decoupling-inspired scheme from Refs.~\cite{Braathen:2019pxr,Braathen:2019zoh}. Interestingly, we found that the DI scheme, while defined in the context of the calculation of the trilinear Higgs coupling, also applies in the present case and ensures the desired renormalization-scale independence as well as apparent and proper decoupling of the BSM contributions to $\Gamma(h\to\gamma\gamma)$.

We have investigated the numerical impact of our new results for two benchmark scenarios inspired by DM phenomenology --- the Higgs resonance scenario with $H$ as a light DM candidate and the heavy Higgs scenario where all inert scalars, including the DM candidate $H$, are heavy. We furthermore imposed for these scenarios state-of-the-art theoretical constraints (in particular perturbative unitarity and vacuum stability) as well as experimental limits from collider and DM searches.

We have shown that the two-loop corrections to $\Gamma(h\to\gamma\gamma)$ can become significant in the presence of large mass splittings (in our case between $m_H\sim\mu_2$ and $m_{H^\pm}=m_A$). While one would not expect to see deviations in \brr~arising from the inert scalars with the present LHC limits, deviations could appear with data from the HL-LHC or from a future $e^+e^-$ Higgs factory. In this regard, the Higgs resonance scenario is most promising and, in this type of scenario, the most of the mass range for $A$ and $H^\pm$ could be ruled out in the near future, considering the expected CMS sensitivity at HL-LHC. It is particularly important to emphasize here that this result requires the inclusion of two-loop BSM corrections to the di-photon decay width. On the other hand, the heavy Higgs scenario would remain out of reach of measurements of the $h\to\gamma\gamma$ decay at the HL-LHC. Prospects are of course better at possible lepton colliders, with (sub)percent level constraints achievable at an electron-positron machine like the ILC --- see for instance Ref.~\cite{Fujii:2017vwa}. Our results illustrate, in both scenarios we considered, that the inclusion of higher-order corrections to $\Gamma(h\to\gamma\gamma)$ is crucial for reliable interpretations of experimental data and expected limits from future colliders. Investigations of the IDM via its effects on Higgs properties should also be considered in complementarity with direct searches for inert scalars at current and future colliders (see $e.g.$ Ref.~\cite{Kalinowski:2020rmb}). Direct collider searches will probe the lower ranges of masses, but it is interesting to note that they will not rule out large deviations in Higgs couplings (as these occur for larger masses of the BSM scalars). Meanwhile, probes via invisible decays of the 125-GeV Higgs boson or via DM direct detection, typically do not significantly constrain scenarios leading to large deviations in \brr~(or in $\kappa_\lambda$), as illustrated by the specific cases considered in this work. On the other hand, while scenarios like the Higgs resonance one could entirely avoid constraints from DM direct detection if the DM detection cross-section is below the neutrino floor, it could potentially be excluded entirely in the future if no BSM deviation is detected in \brr. Our work strengthens the motivation to compute Higgs properties such as its decay width to two photons beyond leading order, in order to reduce theoretical uncertainties and allow reliable comparisons between theoretical predictions and experimental results.


\section*{Acknowledgements}
We thank T.~Katayose for collaboration in the early stages of this project. We thank G. Guedes, M. Spira, and G. Weiglein for helpful discussions, as well as P. Slavich for communication and cross-checks of our work with Ref.~\cite{Degrassi:2023eii}. This work is supported by the Japan Society for the Promotion of Science (JSPS) Grant-in-Aid for Scientific Research on Innovative Areas (No.~22KJ3126 [M.A.]). J.B. acknowledges support by the Deutsche Forschungsgemeinschaft (DFG, German Research Foundation) under Germany's Excellence Strategy -- EXC 2121 ``Quantum Universe'' -- 390833306. J.B. is supported by the DFG Emmy Noether Grant No.\ BR 6995/1-1. This work has been partially funded by the Deutsche Forschungsgemeinschaft (DFG, German Research Foundation) -- 491245950. 
The work of S. K. was supported by the JSPS KAKENHI Grant No. 20H00160, 23K17691, 24KF0060. 
\bigskip

\appendix 

\section{Expressions for the two-loop contributions to the Higgs decay width to two photons, with $\mu_2$ renormalized in the decoupling-inspired scheme}
\label{app:2L_res_IDM_decoup}
We provide in this appendix expressions for the two-loop contributions to the Higgs di-photon decay in the case where the BSM mass parameter $\mu_2$ is renormalized in the decoupling-inspired (DI) scheme defined in Refs~\cite{Braathen:2019pxr,Braathen:2019zoh}.  
The corresponding counterterm, denoted $(\delta^{(1)}\mu_2^2)^\text{\decoup}$, reads
\begin{align}
\label{eq:dmu2_decoup}
    (16\pi^2)(\delta^{(1)}\mu_2^2)^\text{\decoup}=&\ 3 \lambda_2 \mu_2^2\Delta_\text{UV} \\
   - \frac{1}{2}\lambda_2 \mu_2^2 &\bigg[\log\frac{m_H^2}{Q^2} + \log\frac{m_A^2}{Q^2} + 4 \log\frac{m_{H^\pm}^2}{Q^2} - 6\bigg]\,,\nn 
\end{align}
where $Q$ is the renormalization scale. 

While the expressions of $\mathcal{I}^{(2)}_{\mathcal{O}(\lambda_3^2)}$, $\mathcal{I}^{(2)}_{\mathcal{O}((\lambda_4+\lambda_5)^2)}$, and $\mathcal{I}^{(2)}_{\mathcal{O}((\lambda_4-\lambda_5)^2)}$ are unchanged compared to \cref{eq:2Lres_IDM}, the one of $\mathcal{I}^{(2)}_{\mathcal{O}(\lambda_2)}$ is modified and $\mathcal{I}^{(2)}_\text{rem.}$ is simply equal to $\mathcal{I}^{(2)}_\text{ext.-leg.}+\mathcal{I}^{(2)}_\text{VEV}$. 

{\allowdisplaybreaks
\begin{widetext}
\begin{align}
\label{eq:I2_DIscheme}
 \mathcal{I}^{(2)}_{\ord{\lambda_2}}\bigg|^{ (\delta^{(1)}\mu_2^2)^\text{\decoup}}=&-\frac{\lambda_2}{384\pi^2\mhp{2}}\big(m_H^2+m_A^2+4\mhp{2}-6\mu_2^2\big)\,,\nn\\
 \mathcal{I}^{(2)}_\text{rem.}=&\ \mathcal{I}^{(2)}_\text{ext.-leg.}+\mathcal{I}^{(2)}_\text{VEV}
\end{align}
\end{widetext}
}

\section{Expressions for the unrenormalized two-loop contributions to the photon self-energy}
\label{app:2Lphotonself}
For the sake of further applications and cross checks, we provide in this appendix expressions for the unrenormalized two-loop contributions to the photon self-energy in the IDM. These read
\begin{widetext}
\begin{align}
   (16\pi^2)^2\Pi^{(2)}_{\ord{\lambda_3^2}}=&\ \frac{2e^2\mhp{2}}{3v^2} \left(1 - \frac{\mu_2^2}{\mhp{2}}\right)^2\bigg[2\Delta_\text{UV}+1-4\log\frac{\mhp{2}}{Q^2}\bigg]+\ord{\epsilon}\,,\\
  (16\pi^2)^2\Pi^{(2)}_{\ord{(\lambda_4+\lambda_5)^2}}=&\ \frac{e^2}{3v^2} \Bigg\{\mhp{2}\left(1 - \frac{m_H^2}{\mhp{2}}\right)^2\bigg[\Delta_\text{UV}-\log\frac{m_H^2}{Q^2}-\log\frac{\mhp{2}}{Q^2}\bigg]+\frac{m_H^4}{\mhp{2}}+6m_H^2+11\mhp{2} \nn\\
   &\qquad\qquad - \frac{\mhp{2} (17 m_H^2 + \mhp{2})}{m_H^2 - \mhp{2}}\log\frac{m_H^2}{\mhp{2}}\Bigg\}+\ord{\epsilon}\,,\nn\\
   (16\pi^2)^2\Pi^{(2)}_{\ord{(\lambda_4-\lambda_5)^2}}=&\ \frac{e^2}{3v^2} \Bigg\{\mhp{2}\left(1 - \frac{m_A^2}{\mhp{2}}\right)^2\bigg[\Delta_\text{UV}-\log\frac{m_A^2}{Q^2}-\log\frac{\mhp{2}}{Q^2}\bigg]+\frac{m_A^4}{\mhp{2}}+6m_A^2+11\mhp{2} \nn\\
   &\qquad\qquad - \frac{\mhp{2} (17 m_A^2 + \mhp{2})}{m_A^2 - \mhp{2}}\log\frac{m_A^2}{\mhp{2}}\Bigg\}+\ord{\epsilon}\,,\nn\\
   (16\pi^2)^2\Pi^{(2)}_{\ord{\lambda_2}}=&\ \frac{e^2 \lambda_2}{6 \mhp{2}}\bigg\{(m_H^2+m_A^2+ 4 \mhp{2})\bigg[\Delta_\text{UV}+1-2\log\frac{\mhp{2}}{Q^2}\bigg]-m_H^2\log\frac{m_H^2}{\mhp{2}}-m_A^2\log\frac{m_A^2}{\mhp{2}}\Bigg\}+\ord{\epsilon}\nn
\end{align}
\end{widetext}
We emphasize that these expressions contain finite pieces arising from contributions of the form $\epsilon\times 1/\epsilon$, which in the present calculation are found to cancel with terms from subloop renormalization. Employing an $\overline{\text{MS}}$ scheme, we find full agreement between our results and the scalar contributions in eq.~(B.1) of Ref.~\cite{Degrassi:2023eii}.

\section{Leading two-loop corrections to $\lambda_{hhh}$ in the IDM}
\label{app:2L_hhh}
We provide in this appendix expressions for the leading two-loop corrections to the trilinear Higgs coupling $\lambda_{hhh}$ in the IDM --- extending the results of Refs.~\cite{Braathen:2019pxr,Braathen:2019zoh}. As in these references, we obtain the new expressions using the effective-potential approximation, and with a full on-shell renormalization scheme. 

For convenience, we decompose the two-loop corrections to $\lambda_{hhh}$ --- which we denote $\delta^{(2)}\lambda_{hhh}$ --- in three pieces, as
\begin{align}
    \delta^{(2)}\lambda_{hhh}=&\ \delta^{(2)}\lambda_{hhh}\big|_{SS}+\delta^{(2)}\lambda_{hhh}\big|_{SSS}\nn\\\
    &+\delta^{(2)}\lambda_{hhh}\big|_\text{ext-leg+VEV}\,.
\end{align}
In this equation, the three pieces correspond respectively to eight-shaped and sunrise diagrams in the effective potential (see $e.g.$ Ref.~\cite{Martin:2001vx} for a description of two-loop contributions to the effective potential) and to one-loop-squared contributions from external-leg corrections and VEV renormalization. 
We find
{\allowdisplaybreaks
\begin{widetext}
\begin{align}
    \delta^{(2)}\lambda_{hhh}\big|_{SS}^{\mu_2^\text{\decoup}}=\frac{6\lambda_2}{(16\pi^2)^2v^3}&\Bigg[3 \left(m_H^4 - 6 m_H^2 \mu_2^2 - \frac{2 \mu_2^6}{m_H^2}\right)+3 \left(m_A^4 - 6 m_A^2 \mu_2^2 - \frac{2 \mu_2^6}{m_A^2}\right)+4 \left(2m_{H^\pm}^4 - 9 m_{H^\pm}^2 \mu_2^2 - \frac{3 \mu_2^6}{m_{H^\pm}^2}\right)\nn\\
    &+2 m_H^2 m_A^2+4m_H^2m_{H^\pm}^2+4m_A^2m_{H^\pm}^2 + \mu_2^4\left(\frac{m_H^4 + m_A^4 }{m_H^2 m_A^2}+\frac{2 (m_H^4 + m_{H^\pm}^4) }{m_H^2 m_{H^\pm}^2}+\frac{2 (m_A^4 + m_{H^\pm}^4)}{m_A^2 m_{H^\pm}^2}\right)\nn\\
    &+62 \mu_2^4-\frac{2 \mu_2^2}{m_H^2} \left( (m_H^2 - \mu_2^2)^2\log\frac{m_H^2}{m_{H^\pm}^2} - 2  m_H^2 \mu_2^2\log\frac{m_A^2}{m_{H^\pm}^2}\right)\nn\\
    &-\frac{2 \mu_2^2}{m_A^2} \left( (m_A^2 - \mu_2^2)^2\log\frac{m_A^2}{m_{H^\pm}^2} - 2  m_A^2 \mu_2^2\log\frac{m_H^2}{m_{H^\pm}^2}\right)\Bigg]\,.
\end{align}
\begin{align}
    \delta^{(2)}\lambda_{hhh}\big|_{SS}^{\mu_2^\text{PDOS}}=& -\frac{42 m_t^2 (m_H^2 - \mu_2^2)}{(16\pi^2)^2v^5} \bigg[m_H^2\left(1-\frac{\mu_2^2}{m_H^2}\right)^2+m_A^2\left(1-\frac{\mu_2^2}{m_A^2}\right)^2+2\mhp{2}\left(1-\frac{\mu_2^2}{\mhp{2}}\right)^2\bigg]\nn\\
    &-\frac{12 m_H^2 \mu_2^2}{(16\pi^2)^2 v^5}\bigg[m_H^2\left(1-\frac{\mu_2^2}{m_H^2}\right)^2+m_A^2\left(1-\frac{\mu_2^2}{m_A^2}\right)^2+2\mhp{2}\left(1-\frac{\mu_2^2}{\mhp{2}}\right)^2\bigg]\nn\\
    &\quad\times\bigg\{\left(1- \frac{m_A^2}{m_H^2}\right)^3\log\left(1 - \frac{m_H^2}{m_A^2}\right)+2\left(1- \frac{\mhp{2}}{m_H^2}\right)^3\log\left(1 - \frac{m_H^2}{\mhp{2}}\right)\bigg\} \nn\\
    &+\frac{6\mhp{2}(m_H^2-\mu_2^2)}{(16\pi^2)^2v^5}\bigg[m_H^2\left(1-\frac{\mu_2^2}{m_H^2}\right)^2+m_A^2\left(1-\frac{\mu_2^2}{m_A^2}\right)^2+2\mhp{2}\left(1-\frac{\mu_2^2}{\mhp{2}}\right)^2\bigg] \nn\\
    &\quad\times\bigg\{\frac{m_H^2}{m_H^2-\mhp{2}}\log\frac{m_H^2}{\mhp{2}}+\frac{m_A^2}{m_A^2-\mhp{2}}\log\frac{m_A^2}{\mhp{2}}\bigg\}  \nn\\
    &-\frac{m_H^2 - \mu_2^2}{(16\pi^2)^2m_H^2 m_A^2 \mhp{2} v^5}\bigg[m_H^2\left(1-\frac{\mu_2^2}{m_H^2}\right)^2+m_A^2\left(1-\frac{\mu_2^2}{m_A^2}\right)^2+2\mhp{2}\left(1-\frac{\mu_2^2}{\mhp{2}}\right)^2\bigg] \nn\\
    &\quad\times\bigg\{12 m_A^6 \mhp{2} - 23 m_H^2m_A^4\mhp{2} - 2 m_H^2 \mhp{2} \mu_2^4 + m_A^2 \mhp{2}\big(85 m_H^4 + 24 \mhp{4} + 46 \mu_2^4\big) \nn\\
    &\quad\quad- 2 m_H^2m_A^2 \big(23 \mhp{4} + 40 \mhp{2} \mu_2^2 + 2 \mu_2^4\big)\bigg\}\nn\\
    &+\frac{12\lambda_2}{(16\pi^2)^2v^3}\bigg[m_A^4 + 2 \mhp{4} - 4 \mu_2^2\mhp{2} + 3 \mu_2^4 - \frac{m_H^2 \mu_2^4}{m_A^2} - m_A^2 \big(m_H^2 + 2 \mu_2^2\big)\nn\\
    &\qquad\qquad\qquad- \frac{ 2 m_H^2 (\mhp{4} - 3 \mhp{2} \mu_2^2 + \mu_2^4)}{\mhp{2}} + \mu_2^2m_A^2\left(1-\frac{\mu_2^2}{m_A^2}\right)^2\log\frac{m_H^2}{m_A^2}\nn\\
    &\qquad\qquad\qquad+2\mu_2^2\mhp{2}\left(1-\frac{\mu_2^2}{\mhp{2}}\right)^2\log\frac{m_H^2}{\mhp{2}} \bigg]\,.
\end{align}
\begin{align}
    \delta^{(2)}\lambda_{hhh}\big|_{SSS}=\frac{4}{(16\pi^2)^2v^5}&\Bigg\{12 m_H^6\left(1 - \frac{\mu_2^2}{m_H^2}\right)^4+12 m_A^6\left(1 - \frac{\mu_2^2}{m_A^2}\right)^4+24 m_{H^\pm}^6\left(1 - \frac{\mu_2^2}{m_{H^\pm}^2}\right)^4\\
    &+(m_H^2-m_A^2)^2\bigg[\mu_2^6\left(\frac{1}{m_H^4}+\frac{1}{m_A^4}\right)+3(m_H^2+m_A^2)\nn\\
    &\hspace{3cm}+\frac{\mu_2^2}{m_H^2m_A^2}\bigg(3\mu_2^2(m_H^2+m_A^2)-12m_H^2m_A^2-2\mu_2^4\bigg)\bigg]\nn\\
    &+(m_H^2 - m_A^2)^3 \mu_2^2\bigg[\left(\frac{3}{m_H^2} - \frac{\mu_2^4}{m_H^6}\right)\log\left(1-\frac{m_H^2}{m_A^2}\right)-\left(\frac{3}{m_A^2} - \frac{\mu_2^4}{m_A^6}\right)\log\left(\frac{m_A^2}{m_H^2}-1\right)\bigg]\nn\\
    &+2(m_H^2-m_{H^\pm}^2)^2\bigg[\mu_2^6\left(\frac{1}{m_H^4}+\frac{1}{m_{H^\pm}^4}\right)+3(m_H^2+m_{H^\pm}^2)\nn\\
    &\hspace{3cm}+\frac{\mu_2^2}{m_H^2m_{H^\pm}^2}\bigg(3\mu_2^2(m_H^2+m_{H^\pm}^2)-12m_H^2m_{H^\pm}^2-2\mu_2^4\bigg)\bigg]\nn\\
    &+2(m_H^2 - m_{H^\pm}^2)^3 \bigg[\left(\frac{3\mu_2^2}{m_H^2} - \frac{\mu_2^6}{m_H^6}\right)\log\left(1-\frac{m_H^2}{m_{H^\pm}^2}\right)-\left(\frac{3\mu_2^2}{m_{H^\pm}^2} - \frac{\mu_2^6}{m_{H^\pm}^6}\right)\log\left(\frac{m_{H^\pm}^2}{m_H^2}-1\right)\bigg]\nn\\
    &+2(m_A^2-m_{H^\pm}^2)^2\bigg[\mu_2^6\left(\frac{1}{m_A^4}+\frac{1}{m_{H^\pm}^4}\right)+3(m_A^2+m_{H^\pm}^2)\nn\\
    &\hspace{3cm}+\frac{\mu_2^2}{m_A^2m_{H^\pm}^2}\bigg(3\mu_2^2(m_A^2+m_{H^\pm}^2)-12m_A^2m_{H^\pm}^2-2\mu_2^4\bigg)\bigg]\nn\\
    &+2(m_A^2 - m_{H^\pm}^2)^3 \bigg[\left(\frac{3\mu_2^2}{m_A^2} - \frac{\mu_2^6}{m_A^6}\right)\log\left(1-\frac{m_A^2}{m_{H^\pm}^2}\right)-\left(\frac{3\mu_2^2}{m_{H^\pm}^2} - \frac{\mu_2^6}{m_{H^\pm}^6}\right)\log\left(\frac{m_{H^\pm}^2}{m_A^2}-1\right)\bigg]\Bigg\}\,.\nn
\end{align}
\begin{align}
    \delta^{(2)}\lambda_{hhh}\big|_\text{ext-leg+VEV}=\frac{1}{(16\pi^2)^2v^5}&\Bigg\{42m_t^2\left[m_H^4\left(1-\frac{\mu_2^2}{m_H^2}\right)^3+m_A^4\left(1-\frac{\mu_2^2}{m_A^2}\right)^3+2m_{H^\pm}^4\left(1-\frac{\mu_2^2}{m_{H^\pm}^2}\right)^3\right]\nn\\
    &+24m_t^4\left[m_H^2\left(1-\frac{\mu_2^2}{m_H^2}\right)^2+m_A^2\left(1-\frac{\mu_2^2}{m_A^2}\right)^2+2m_{H^\pm}^2\left(1-\frac{\mu_2^2}{m_{H^\pm}^2}\right)^2\right]\nn\\
    &-2\left[m_H^4\left(1-\frac{\mu_2^2}{m_H^2}\right)^3+m_A^4\left(1-\frac{\mu_2^2}{m_A^2}\right)^3+2m_{H^\pm}^4\left(1-\frac{\mu_2^2}{m_{H^\pm}^2}\right)^3\right]\nn\\
    &\hspace{1cm}\times\left[m_H^2\left(1-\frac{\mu_2^2}{m_H^2}\right)^2+m_A^2\left(1-\frac{\mu_2^2}{m_A^2}\right)^2+2m_{H^\pm}^2\left(1-\frac{\mu_2^2}{m_{H^\pm}^2}\right)^2\right]\Bigg\}\,.
\end{align}
\end{widetext}
}
These expressions are manifestly independent of the renormalization scale, as expected from our choice of renormalization prescription. Moreover, we have verified: \textit{(i)} that these BSM effects decouple properly in the limit $\mu_2\to\infty$, and \textit{(ii)} that in the limit $\mu_2,m_H\to 0$ we recover equation (V.22) of Ref.~\cite{Braathen:2019zoh}. Finally, we note that the expression for $\delta^{(2)}\lambda_{hhh}\big|_{SSS}$ is provided here for the mass hierarchy $m_H<m_A<m_{H^\pm}$, but it can be adapted for any desired hierarchy by selecting the appropriate branch of the logarithms.  

\bibliography{biblio}

\end{document}